\documentclass[11pt, usenames]{amsart}
\usepackage{xr}
\makeatletter
\newcommand*{\addFileDependency}[1]{% argument=file name and extension
  \typeout{(#1)}% latexmk will find this if $recorder=0 (however, in that case, it will ignore #1 if it is a .aux or .pdf file etc and it exists! if it doesn't exist, it will appear in the list of dependents regardless)
  \@addtofilelist{#1}% if you want it to appear in \listfiles, not really necessary and latexmk doesn't use this
  \IfFileExists{#1}{}{\typeout{No file #1.}}% latexmk will find this message if #1 doesn't exist (yet)
}
\makeatother

\newcommand*{\myexternaldocument}[1]{%
    \externaldocument{#1}%
    \addFileDependency{#1.tex}%
    \addFileDependency{#1.aux}%
}
\myexternaldocument{JASA-supp-v1}

\usepackage{amsmath}
\usepackage{tikz}

\usetikzlibrary{topaths, calc, trees, positioning, arrows, shapes, shapes.multipart, shadows, matrix, decorations.pathreplacing, decorations.pathmorphing,backgrounds}

\usepackage{amsfonts,amssymb,amsthm, txfonts, pxfonts,amscd}
\usepackage[stretch=10]{microtype}
\def\struckint{\mathop{%
\def\mathpalette##1##2{\mathchoice{##1\displaystyle##2}%
 {##1\textstyle##2}{##1\scriptstyle##2}{##1\scriptscriptstyle##2}}%
\mathpalette
{\vbox\bgroup\baselineskip0pt\lineskiplimit-1000pt\lineskip-1000pt
\halign\bgroup\hfill$}
{##$\hfill\cr{\intop}\cr\diagup\cr\egroup\egroup}%
}\limits}
\usepackage{booktabs,caption,fixltx2e}
\usepackage{multirow}
\usepackage[flushleft]{threeparttable}
\usepackage{hyperref}
\usepackage{standalone}
\usepackage{layouts}
\usepackage{setspace}
\usepackage{xcolor}
\def\pr{\mathop{\text{pr}}\nolimits}

\def\Nb{\mathop{\mathbb{N}_{}}\nolimits}

\def\fin{\mathop{\text{fin}}\nolimits}
\def\pr{\mathop{\text{pr}}\nolimits}
\def\E{\mathcal{E}}

\def\H{\mathcal{H}}

\def\equalinlaw{=_{\mathcal{D}}}

\newtheorem{thm}{Theorem}[section]
\newtheorem{lemma}[thm]{Lemma}

\newtheorem{defn}[thm]{Definition}
\newtheorem{example}[thm]{Example}

\newtheorem{rmk}[thm]{Remark}%\endlocaldefs

\makeatletter

\def\S{\mathcal{S}}
\def\R{\mathcal{R}}
\def\H{\mathcal{H}}
\def\Y{\mathcal{Y}}
\def\E{\mathcal{E}}
\def\EE{\mathbb{E}}
\def\P{\mathcal{P}}
\def\ybf{{\bf y}}
\def\Ybf{{\bf Y}}
\def\Nat{\mathbb{N}}
\def\I{\mathcal{I}}
\def\given{\, | \,}
\def\dotminussym#1#2{%
  \setbox0=\hbox{$\m@th#1-$}%
  \kern.5\wd0%
  \hbox to 0pt{\hss\hbox{$\m@th#1-$}\hss}%
  \raise.6\ht0\hbox to 0pt{\hss$\m@th#1.$\hss}%
  \kern.5\wd0}

\mathchardef\mhyphen="2D

\definecolor{WowColor}{rgb}{.75,0,.75}
\definecolor{SubtleColor}{rgb}{0,0,.50}

\begin{document}

\title{Hierarchical network models for structured exchangeable
  interaction processes}
\author{Walter Dempsey and Brandon Oselio and Alfred Hero}
%\thanks{R code and links to datasets publicly available at https://github.com/wdempsey/e2}

\date{\today}

\begin{abstract}
Network data often arises via a series of \emph{structured
  interactions} among a population of constituent elements. 
E-mail exchanges, for example, have a single sender followed by potentially
multiple receivers. 
Scientific articles, on the other hand, may have multiple subject
areas and multiple authors.
We introduce {\em hierarchical edge exchangeable
models} for the study of these structured interaction networks.
In particular, we introduce the {\em hierarchical vertex components
  model} as a canonical example, which partially pools information via
a latent, shared population-level distribution. 
Theoretical analysis and supporting simulations provide clear model
interpretation, and establish global sparsity and power-law degree
distribution. A computationally tractable Gibbs algorithm is derived.
We demonstrate the model on both the Enron e-mail dataset and an ArXiv dataset,
showing goodness of fit of the model via posterior predictive validation.
% Moreover, interaction data often include temporal information; 
% We introduce \emph{dynamic, hierarchical edge exchangeable model} 
% for temporal structured interaction data. An inferential algorithm is
% derived and the dynamic Enron model is fit to the time-stamped Enron dataset.
\end{abstract}


\maketitle

\section{Introduction}\label{section:introduction}

Modern statistical network analysis focuses on the study of large,
complex networks that can emerge in diverse fields, including social,
biological, and physical systems~\cite{Barabasi2016, Newman2010,
  Estrada2012, Latora2017, FienbergAiroldiSurvey}. The expanding scope
of network analysis has led to an increase in the need for statistical
models and inferential tools that can handle the increasing complexity
of network data structures. 
In this paper, we focus on network data arising from sequences of
interactions. Network data arising in this manner would benefit from a
 framework built upon the interaction as the 
statistical unit~\cite{McCullaghAOS} rather than upon the constituent
elements within each interaction as the statistical units. %  -- replacing
% the standard vertex-centric perspective with a relational one.
\emph{Edge-exchangeable models}~\cite{CraneJASA, CraneSTS} are built
specifically to analyze datasets containing these complex
interactions. 

While Crane and Dempsey (2017)~\cite{CraneJASA} provide a framework
for statistical analysis of interaction data, the proposed \emph{Hollywood
  model} only captures basic global features. Specifically, the
Hollywood model's asymptotic behavior reflects the empirical properties of
sparsity and power law degree distributions observed in real-world
network data, which are not as well reflected in classic statistical network
models such as the ERGMs~\cite{Wasserman1996}, graphon
models~\cite{Airold2013}, and stochastic blockmodels
(SBMs)~\cite{Holland1983}. While edge exchangeability is attractive as
a theoretical framework, the set of current edge exchangeable models
is inadequate to handle the structural complexity of modern network data.

The edge exchangeable model proposed in this paper is motivated by an
important fact: most common complex networks
constructed from interaction data are \emph{structured}.  A
phone-call interaction, for example, takes the form of a sender and
receiver pair.  E-mail correspondence generalizes this type of
interaction to one sender but potentially multiple receivers with different
attributes like ``To,'' ``Cc,'' and ``Bcc''. 
This paper makes a substantial push forward by constructing 
hierarchical models that reflect this common structure of interaction
data, hereafter referred to as \emph{structured interaction data}.
The model overlays local behavior (e.g., per sender) with global
information by partial pooling through a shared global, latent 
distribution.
Simulation and theoretical analysis confirm that the proposed
hierarchical model can achieve simultaneously varying local power-law 
degree per sender and global power-law degree distribution.

\subsection{Relevant prior work on interaction data}

Interaction data often arises in settings where 
communications amongst a set of constituent elements
over a specific time period are recorded~\cite{Tyler2005, Eagle2006}.
Examples are numerous and include: 
authorship and co-sponsoring of legislation~\cite{Fowler2006, Signorelli2018},
sending and receiving e-mails~\cite{Tyler2005, Cohen2009},  
posting and responding on a community forum~\cite{coursera-iri2014},
and traceroute~\cite{luckie2008}.
In each case, the interaction (edge) is the statistical unit to be modeled, as
contrasted with the subjects (nodes) of the interactions considered in other work~\cite{goldenberg2010survey}. 
See~\cite{CraneSTS, CraneJASA} for further discussion of the advantages of defining
interactions as the statistical units.

% \textcolor{red}{Do we need this paragraph? Suggest delete}
% \textcolor{red}{Keeping to just show examples of interaction data modeling}
The literature contains several papers focused on statistical modeling
of interaction data. Perry and Wolfe (2013)~\cite{PerryWolfe2013}
construct a Cox proportional intensity model~\cite{Cox1972}.  
Butts (2008)~\cite{Butts2008} considered likelihood-based inference
using a variant of the proportional intensity model to capture
interaction behavior in social settings.
Crane and Dempsey (2017)~\cite{CraneJASA} consider non-hierarchical models 
for interaction data.  They introduce the notion of edge
exchangeable network models and explore its basic statistical
properties. In particular, they show that edge exchangeable models
allow for sparse structure and power law degree distributions, widely
observed empirical behaviors that cannot be handled by conventional
approaches. 

An alternative approach emerges out of the recent work of 
Caron and Fox (2017) \cite{CaronFox2017}, who construct random graphs from
point processes on $\mathbb{R}_{+}^2 = [0,\infty)\times[0,\infty)$.
The random graph is characterized by an object called a
\emph{graphex}~\cite{VeitchRoy2015}.
Random graph models generated by this procedure
can incorporate sparse, power law behavior 
into a well-defined population model.
Finite random graphs can be obtained via 
a thresholding operation, termed \emph{p-sampling}~\cite{VeitchRoy2016}.
Such random graphs are vertex exchangeable in that they are built from 
exchangeable point processes. 
In this setting, exchangeability is a consequence of projectivity
rather than the simple structured interaction data sampling scheme proposed in
this paper.
See the contributed discussion to the paper by Caron and Fox
(2017)~\cite{CaronFox2017}, in particular contributions by
Bharath~\cite{BharathJRSSB-discussion} and
Crane~\cite{CraneJRSSB-discussion}, for further discussion.

\subsection{Outline and main contributions}

The main contributions of this paper are as follows: 
\begin{enumerate}
\item We start by formally defining structured interaction data in
  Definition~\ref{defn:data}.
  We then define exchangeable structured interaction processes in
  Definition~\ref{defn:exchproc}.
\item We prove a representation theorem for these exchangeable
  processes in Theorem~\ref{thm:repthm}; we then define, in
  section~\ref{section:hvcm}, \emph{hierarchical vertex components
    models} (HVCM) -- a subfamily of exchangeable processes that
  capture important interaction dynamics. 
\item A particular computationally tractable HVCM is introduced in
  section~\ref{section:emailmodel} and an efficient Gibbs sampling
  inferential algorithm is derived in section~\ref{section:estimation}. 
\item We establish basic statistical properties in
  section~\ref{section:properties}.  In particular, we provide theoretical
  guarantees of sparsity and power law for the chosen HVCM -- two
  important empirical properties of network data.
\item We demonstrate this HVCM on both the Enron e-mail
  dataset and ArXiv dataset in
  section~\ref{section:hierarchical_application}. In particular, we 
  show how the HVCM can be used to perform goodness of fit checks for models of network
  data via posterior predictive checks, an often under-emphasized aspect of  
  statistical network modeling. 
\end{enumerate}
Overall, this paper presents a statistically rigorous, principled 
hierarchical modeling framework for handling complex structured
interaction data.

% The canonical hierarchical vertex components model introduced in
% section~\ref{section:emailmodel} arises as a special case.  We then
% establish the basic statistical properties in
% section~\ref{section:properties}
% and present an inference procedure in
% section~\ref{section:estimation}. We apply this model to the Enron
% email network ,
% establishing posterior predictive checks.
% Section~\ref{section:temporaldata} moves on to temporal interaction
% data, introducing the temporal Enron model. \textcolor{red}{delete next?}
% An inference procedure is proposed and applied to the Enron data in
% sections~\ref{subsection:inf_tempEnron}
% and~\ref{subsection:casestudy_tempEnron} respectively.

\section{Structured interaction data}\label{section:networkdata}

\noindent We start by defining structured interaction data,
illustrating with a sequence of concrete examples of increasing
complexity. 

\begin{defn}[Structured interaction data]\label{defn:data}
\normalfont
Let~$\mathcal{P}$ denote a set of constituent elements.
Then for a set $\mathcal{P}$, we write $\fin(\mathcal{P})$ to denote the
set of all finite multisets of $\mathcal{P}$.
A {\em structured interaction process} for
an ordered sequence of sets~$(\mathcal{P}_1,\ldots, \mathcal{P}_k)$ is a correspondence
$\I: I \to \fin(\mathcal{P}_1) \times \ldots \times \fin(\mathcal{P}_k)$ between a set $I$ indexing
interactions and the ordered sequence of finite multisets
of~$(\mathcal{P}_1,\ldots, \mathcal{P}_k)$.
\end{defn}

\begin{rmk}[Difference from interaction data] \normalfont
In~\cite{CraneJASA}, an interaction process is defined as a
correspondence~$\I: I \to \fin (\P)$ where $\P$ is a single
population.
Structured interaction data, instead, consists of a series of finite
multisets, and does not require each set of constituent elements to be
equivalent. That is, each population~$\mathcal{P}_k$ may contain
different types of elements. This flexibility will allow us to
introduce hierarchical structure into the exchangeable model.
\end{rmk}

\noindent Finally, let~$\fin_k (\mathcal{P})$ denote the multisets of
size~$k$, so that $\fin(\mathcal{P})$ is the disjoint
union~$\cup_{k=1}^\infty \fin_{k} (\mathcal{P})$.

\begin{example}[Phone-calls]
\normalfont
Assume~$\mathcal{P}_k$ are all equivalent and
let~$\mathcal{P}_k=:\Nat$ be a countably infinite population.  A
phone-call can be represented as an ordered pair of ``sender'' and
``receiver'' drawn from~$\Nat$. 
Therefore, a phone-call interaction process is a
correspondence~$\mathcal{I}: I \to \fin_{1} (\mathbb{N}) \times
\fin_{1} (\mathbb{N})$. For instance,~$I(1)= ( \{ a \} , \{ b \} )$ is
a phone-call from sender~$a$ to receiver~$b$, both in
population~$\Nat$. This is distinct from $( \{ b \}, \{ a \})$ where
sender and receiver roles are reversed.
\end{example}

% \begin{example}[Citation]
% \normalfont
% Assume~$\mathcal{P}_k$ are equivalent and let $\P_k = \Nat$ be a
% countably infinite population. A citation is a pair of an article and
% a set of articles that the first article cites that the first article
% cites. Thus, a citation process is a correspondence
% $\mathcal{I}: I \to \fin_1 (\mathbb{N}) \times \fin (\mathbb{N})$.
% For instance,~$I(1)= (\{a\}, \{b ,c \})$ is an article~$a$ that cites
% articles~$b$ and~$c$.
% \end{example}

\begin{figure}[!th]
  \centering
  \includegraphics[width=3.5in]{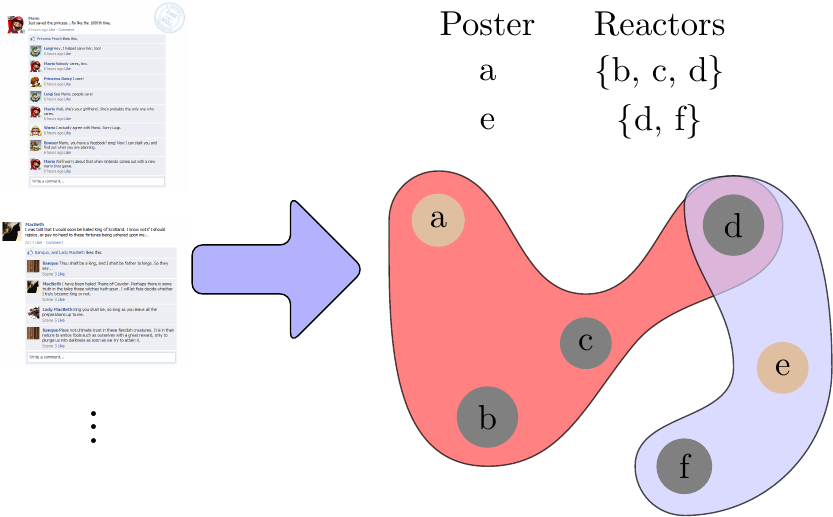} 
  \caption{Example of network data from Facebook posts. The post
    process is a correspondence~$\mathcal{I}:I \to \fin_1 (\Nb) \times
    \fin (\Nb)$. Here~$I = \Nb$, and the first post~$\mathcal{I}(1) =
    \{ \{a\}, \{b,c,d\}\}$ represents user~$a$ posting to the forum
    and $b,c,d$ reacting to the post. The second post~$\mathcal{I}(2)
    = \{\{e\}, \{d,f\}\}$ represents user~$e$ posting to the forum and
    $d,f$ reacting.  User~$d$ reacts to both posts.    
}
  \label{fig:concept}
\end{figure}

\begin{example}[E-mails] \label{ex:emails}
\normalfont
Assume~$\mathcal{P}_k$ are all equivalent and
let~$\mathcal{P}_k = \mathbb{N}$ be a countably infinite population.  
An e-mail can be represented as the ordered sequence of sets: sender,
receivers. 
%, carbon copy (cc) receivers, and blind carbon copy (bcc) receivers.
% Define~$\widetilde{\fin} (\mathcal{P})$ to be the union of~$\fin
% (\mathcal{P})$ and the empty set~$\emptyset$.
Then an e-mail interaction process is a correspondence
$\mathcal{I}: I \to \fin_1 (\mathbb{N}) \times \fin (\mathbb{N})$.
For instance,~$I(1)= ( \{ a \}, \{ b, c \})$ is an e-mail from sender~$a$ to
receivers~$b$ and $c$. This is distinct from $( \{ b \}, \{ a, c\})$
and $(\{c \}, \{ a, b\} )$.
Figure~\ref{fig:concept} is a visualization of a similar structured
interaction dataset formed from Facebook posts (i.e., poster followed
by finite multiset of responders).  
% $\mathcal{I}: I \to \fin_1 (\mathbb{N}) \times \widetilde{\fin} (\mathbb{N})
% \times \widetilde{\fin} (\mathbb{N}) \times \widetilde{\fin}
% (\mathbb{N})$, where each e-mail requires at least one receiver, cc receiver, or bcc receiver.
% For instance,~$I(1)= (a, \{ b, c \}, \emptyset, \emptyset)$ is an e-mail from sender~$a$ to
% receivers~$b$ and $c$ with no cc nor bcc receivers.
% This is distinct from $(a, \emptyset, \{ b, c\}, \emptyset)$ and
% $(a, \emptyset, \emptyset, \{ b, c\} )$ where the receivers become cc
% receivers and bcc receivers respectively.
\end{example}

\begin{example}[Scientific articles]
\normalfont
Consider summarizing a scientific article by its (1) list of subject
areas and (2) list of authors. 
Then the scientific article process is a correspondence
$\mathcal{I}: I \to \fin (\P_1) \times \fin (\P_2)$.
For instance,~$I(1)= (\{ a, b \}, \{c, d \})$ is an article with subject
areas~$a$ and~$b$ and authors~$c$ and $d$.
Here,~$\P_1$ and~$\P_2$ are distinct populations.
\end{example}

\begin{example}[Movies] \normalfont
% Assume~$\mathcal{P}_k$ are not equivalent, but potentially overlapping
% countably infinite populations.
Consider summarizing a movie by its (1) genre, (2) list of
producers, (3) director, and (4) list of actors.
Of course, there is overlap in certain populations, as producers can
be directors, directors can be actors, but none are a
genre (unless Scorsese, Spielberg, or Tarantino are considered genres
unto themselves).
Then the movie process is a correspondence
$\mathcal{I}: I \to \fin_1 (\mathbb{N}) \times \fin (\mathbb{N})
\times \fin_1 (\mathbb{N}) \times \fin (\mathbb{N})$.
For instance,~$I(1)= (\{a\}, \{ b, c \}, \{ d \}, \{d, e, f\})$ is a movie
with genre~$a$, producers~$b$~and~$c$, director~$d$, and actors~$
d$, $e$, and $f$. Note, in this example, the director is also one of the
actors.
\end{example}

% \begin{example}[Traceroute] \normalfont
% Assume~$\mathcal{P}_k$ are all equivalent and
% let~$\mathcal{P}_k = \mathbb{N}$ be a countably infinite population.
% A traceroute is the path traversed by a message sent from two IP
% addresses over the Internet -- from a \emph{source}~$s$ to a
% \emph{target}~$t$.
% The intermediary nodes are ordered~$(n_{1}, \ldots, n_{k})$ in that
% the path goes from $s$ to $n_1$ to $n_2$ and so on until it reaches
% the target~$t$. This set of nodes can be empty as a path can go
% directly from source to target.
% Then, a traceroute interaction process is a correspondence
% $\mathcal{I}: I \to \cup_{k=0}^\infty \fin_1 (\mathbb{N}) \times \otimes_{j=1}^{k}
% \fin_1 (\mathbb{N}) \times \fin_1 (\mathbb{N})$. 
% For instance,~$I(1)= ( \{s\}, \{ n_1 \}, \{ n_2 \}, \{t\})$.
% \end{example}

\noindent The above shows Definition~\ref{defn:data} covers a wide
variety of examples from network science. 
% In addition, this model does not remove or
% summarize the collected data - as most network models do - by reducing
% the interactions to an undirected or directed graph with interaction
% tuples. 
Next, we construct interaction-labeled networks and define
exchangeable structured interaction processes. % using Definition
%\ref{defn:data}. % Construction and definitions require minor
% modifications for particular examples of interaction data, but these
% additional details are omitted.   

\begin{rmk}[Covariates] \normalfont
In this paper, we focus on the study of structured
interaction processes in Definition \ref{defn:data} with no additional
information, such as covariates. 
% In Section~\ref{section:dynamic} we incorporate temporal information
% (i.e., time-stamps associated with each particular interaction).
% All other covariate information is ignored; 
Incorporating such covariate information is quite difficult;
see~\cite{AiroldiChoi2011, Latouche2015, Mariadassou2010, Sweet2015,
  Tallberg2004, ZhangLevina2015} for examples of incorpating
covariates into network analysis.  
Covariate information can come in two forms: (1) covariate information
on the interaction; and (2) covariate information on constituent
elements.
Examples of (1) include subject line or body text in an e-mail, or
genre and gross movie sales for a movie.  Examples of (2) include
gender, age, job title, or university affiliation of authors of a
scientific article.
Certain interaction covariates can be incorporated into the models
considered in this paper.  For example, in the ArXiV dataset, the
article's subject can be viewed as covariate information on the
interaction.  We show how this can be incorporated as part of the
structured interaction data structure, and therefore accounted for in
the statistical models.
\end{rmk}

\subsection{Interaction-labeled networks}

For the remainder of this paper, we focus on structured interaction
processes of the form~$\I: I \to \fin (\mathcal{P}_1) \times
\fin(\mathcal{P}_2)$. This type of a structured interaction process
captures the phone-call, e-mail, and scientific article examples.
The arguments presented naturally extend to more general structured interaction
processes as given in Definition~\ref{defn:data}.
% See section~\ref{section:discuss} for discussion of the traceroute
% example. \textcolor{red}{delete previous sentence?}
When two populations of constituent elements are equivalent, we
write~$\mathcal{P}_1 \equiv \mathcal{P}_2$.
% sufficiently general and it is easy to see how the below can be
% adjusted for the each example in the previous section.
% if the two populations are the same for~$i \neq j$, e.g., population
% of senders and receivers.
The interaction-labeled network is an equivalence clase constructed from the structured
interaction process by quotienting out the labeling of the constituent
elements. 
% {\color{red}Do we need to define}
%   $\mathcal{P}_i^'$\textcolor{red}{?, i.e. another set of constituent elements, such that}
  % $|\mathcal{P}| = |\mathcal{P}|'$
Let~$\rho_j: \P_j \to \P_j^\prime$ be a bijection
for~$j = 1,2$.
We write~$\rho: \P_1 \times \P_2 \to \P_1^\prime \times \P_2^\prime$
to be the composite bijection obtained by applying $\{ \rho_j
\}_{j=1,2}$ componentwise.
If~$\mathcal{P}_1 \equiv \mathcal{P}_2$, then $\rho_1 \equiv \rho_2$;
that is, bijections among equivalent populations, e.g., the senders and
receivers in an email network, denoted by $ \bar s$ and $\bar r$, respectively, must agree.
Then $\rho$ induces an action on the product space~$\fin
(\mathcal{P}_1) \times \fin (\mathcal{P}_2)$ by the composite map 
% \textcolor{red}{does this have a specific meaning? could we just say
% ``induces a composite map on the product space...''}
\begin{align*}
  ( \bar s , \bar r )
  &= ( \{ s_1, \ldots, s_{k_1} \} , \{r_{1}, \ldots, r_{k_2}\} )
    \in \fin (\P_1) \times \fin (\P_2) \\
  \to \rho \, (\bar s , \bar r)
  &= ( \{ \rho_1 s_1, \ldots, \rho_1 s_{k_1} \} , \{ \rho_2 r_{1}, \ldots, \rho_2
    r_{k_2}\})  \in \fin (\P^\prime_1) \times \fin (\P^\prime_2) 
\end{align*}
Therefore, the bijection~$\rho$ acts on the structured
interaction process via composition~$(\rho \I) (i) = \rho (
\I (i) ), i \in \mathbb{N}$.
The structured interaction-labeled network is then the equivalence
class constructed from the structured interaction network by
quotienting out over bijections~$\rho$:
\begin{equation} \label{eq:cong}
  \ybf_{\I} = \bigcup_{ \substack{\# \P_j^\prime = \# \P_j \\ j =1,2}} \left \{ \I^\prime : I \to
    \fin (\P^\prime_1) \times \fin (P^\prime_2) :
    \rho \I = \I^\prime \text{ for some bijection } 
    \rho : \P_1 \times \P_2 \to \P^\prime_1 \times \P_2^\prime  \right
  \},
\end{equation}
where~$\# \P_j$ is the cardinality of the population.
Note we have only quotiented out labels for constituent elements,
so the object~$\ybf_{\I}$ still has uniquely labeled interactions. For
simplicity, we write~$\ybf$ and leave the subscript~$\I$ implicit.

In the remainder of the paper, we assume the index set~$I$ is
countably infinite and replace it by~$\mathbb{N}$.
For any $S \subset \mathbb{N}$, we define the \emph{restriction} of 
$\mathcal{I}:\mathbb{N} \to \fin (\P_1) \times \fin(\P_2)$ to the subset~$S \subset \mathbb{N}$ 
by~$\mathcal{I}|_{S}$.  This restricted interaction process induces a
restriction to $S$ of the interaction-labeled network.  We
write~$\ybf_{S}$ to denote the interaction-labeled network
associated with the restricted process~$\mathcal{I}|_{S}$.
For~$S = [n] := \{1,\ldots, n\}$, we simply write~$\mathcal{I}_n$ to denote the
restricted structured interaction process and $\ybf_n$ to denote
the corresponding structured interaction network. 

\section{Structured interaction exchangeable models}\label{section:markove2}

% Here we briefly overview a special-case of the de Finetti-type
% representation theorem of edge exchangeable networks for structured
% interaction data to motivate our modeling approach. We start by
% defining the relabeling of the interaction network.  Note this is 
% different from relabeling of the constituent elements via the
% bijection $\rho$. 

Let~$\ybf$ denote the interaction-labeled network constructed
from the structured interaction process~$\I : \mathbb{N} \to
\fin (\P_1) \times \fin (\P_2)$.  Then for any finite permutation~$\sigma :
\mathbb{N} \to \mathbb{N}$, let~$\I^{\sigma}$ denote the relabeled structured
interaction process defined by ~$\I^{\sigma} (i) = \I (\sigma^{-1}
(i)), i \in \mathbb{N}$. 
% \textcolor{red}{why inverse here?}
% \textcolore{red}{ So it is inverse because if \sigma(1) = 2,
% \sigma(2) = 3 and \sigma (3) = 1 (all other stay same).
% Then the first interaction in I^\sigma is I(3) = I(\sigma^{-1}
% (1)). Annyoing but true}
Then $\ybf^{\sigma}$ denotes the corresponding interaction
labeled network constructed from~$\I^\sigma$. 
Note that the choice of representative from the equivalence class does
not matter.
The above relabeling by permutation~$\sigma$ is not to be confused
with the relabeling in the previous section by the bijection~$\rho$.
The bijection~$\rho$ relabels the constituent elements, and is
used to construct the equivalence class defining the interaction-labeled network (i.e., the
equivalence class). 
The permutation~$\sigma$ reorders the interaction process, and
therefore relabels the interaction-labeled network. 

In the remainder of this paper, we write~$\Ybf$
% \textcolor{red}{Why don't we just
%   stick to $\mathcal{E}$? moving to $\Ybf$ cause then it's clear that
%   this is the response. Thoughts?} 
to denote a random
interaction-labeled network. 
%We write~$\EN$ to denote the space of interaction exchangeable networks.
We assume the interactions are labeled in the countably infinite
set~$\Nb$.
Interaction exchangeability is characterized by the property that the 
labeling of the interactions (not the constituent elements) 
is arbitrary.
We now define exchangeable structured interaction networks.

\begin{defn}[Exchangeable structured interaction network process] 
\label{defn:exchproc}\normalfont
The structured interaction-labeled network~$\Ybf$ is exchangeable if
$\Ybf^{\sigma} \equalinlaw \Ybf$ for all permutations $\sigma : \Nb \to
\Nb$, where $\equalinlaw$ denotes equality in distribution.
\end{defn}

Next, we provide a representation theorem for structured interaction
processes.  We focus on the setting where each interaction~$(\bar s,
\bar r)$ is either never observed or observed infinitely often. This
is commonly referred to as the ``blip-free''
setting~\cite{CraneDempsey}, where blips refer to interactions $(\bar
s, \bar r)$ that are observed once.   We first define the $\fin (\P_1)
\times \fin (\P_2)$-simplex
\[
\mathcal{F} = \left \{ (f_{(\bar s, \bar r)})_{(\bar s,\bar r) \in \fin (\P_1)
    \times \fin (\P_2)} \quad \text{and} \quad \sum_{(\bar s,\bar r)
    \in \fin (\P_1) \times \fin (\P_2)} f_{(\bar s,\bar r)} = 1  \right
\}
\]
where~$(\bar s, \bar r) := (\{s_1, \ldots, s_{k_1} \}, \{r_1, \ldots,
r_{k_2}\})$ for $s_1,\ldots, s_{k_1} \in \P_1$ and $r_1,\ldots, r_{k_2} \in \P_2$.
Let~$\phi$ be a probability measure on the simplex and define~$f \sim \phi$
to be a random variable drawn from this measure.
Then, given~$f \in \mathcal{F}$, let the sequence of
interactions~$\I(1),\I(2), \ldots$ be generated according to
\begin{equation}
\label{eq:genXs}
\pr \left (\I(i) = \left( \{ s_1, \ldots, s_{k_1} \}, \{r_1, \ldots,
  r_{k_2} \} \right) \given f  \right) = f_{(\bar s,\bar r)}.
\end{equation}
Then, given~$\I$, set~$\Ybf = \ybf_{\I}$. Theorem~\ref{thm:repthm}
states that all blip-free structured interaction exchangeable networks 
can be generated in this manner. The proof can be found in
Section~\ref{section:rep} of the supplementary materials.

\begin{thm}[Blip-free representation theorem]
\label{thm:repthm} \normalfont
Let~$\Ybf$ be a structured interaction exchangeable network
that is blip-free with probability 1.  Then there exists a probability
measure~$\phi$ on $\mathcal{F}$ such that $\Ybf \sim \epsilon_{\phi}$,
where 
\[
\epsilon_{\phi} (\cdot) = \int_{\mathcal{F}} \epsilon_f (\cdot) \phi (df).
\]
% So, every blip-free exchangeable structured interaction network $\Ybf$
% can be generated via~$f \sim \phi$ and then $\Ybf =
% \ybf_{\I}$ with structured interaction process $\I: \mathbb{N} \to
% \fin (\P_1) \times \fin (\P_2)$ from~\eqref{eq:genXs}.
\end{thm}

% This theorem is a consequence of a more general result found
% in~\cite{CraneDempsey2016relational}, which allows for more general structures
% in the interaction process.

\subsection{Hierarchical vertex components model}
\label{section:hvcm}

Via Theorem~\ref{thm:repthm}, we can construct a particular family of
interaction exchangeable random networks as follows.  First,
choose a distribution of senders, $f^\prime=(f_s )_{s \in \P_1}$, in the simplex
\[
  \mathcal{F}_{1}:=\left\{(f_s)_{s\in\P_1}:
    f_s\geq0\quad\text{and}\quad\sum_{s\in\P_1}f_s=1\right\}.
\]
Next, choose a second element of~$\mathcal{F}_1$, which we denote~$w$.  
% Each element of the simplex corresponds to a distribution over the first
% component~$\fin_1 (\P_1)$.
Finally, for each~$s \in \P_1$, construct a conditional distribution over the
receivers, i.e., the second
component~$\fin (\P_2)$.  That is, for every~$s \in \P_1$, we
choose $f^{\prime \prime}_s=(f_{r \given s})_{r \in \P_2}$ in the simplex
\[
  \mathcal{F}_2=\left\{(f_{r} )_{r\in \P_2}:
    f_{r}\geq0\quad\text{and}\quad\sum_{r\in \P_2} f_{r}=1\right\}.
\]
We combine these distributions to form $f\in\mathcal{F}_1 \times
\mathcal{F}_1 \times (\otimes_{s\in\P_1} \mathcal{F}_2)$, which
determines a distribution on the space~$\fin_1 (\P_1) \times
\fin(\P_2)$ by
\begin{equation}
  \pr \left(E = (\bar s, \bar r) \mid f \right) = \nu_{k_1} \left[
  \prod_{i=1}^{k_1} f_{s_i} \right] \cdot
  \frac{ \sum_{i=1}^{k_1} w_{s_i} \cdot \nu^{(s_i)}_{k_2}  \left[
      \prod_{j=1}^{k_2} f_{r_j \given s_i} \right] }{\sum_{i=1}^{k_1} w_{s_i} }, \label{eq:paintbox}
\end{equation}
where~$\nu_{l} \geq 0$, $\nu^{(s)}_l \geq 0$, $\sum_{l=1}^\infty
\nu_{l} = 1$, and $\sum_{l=1}^\infty \nu_{l}^{(s)} = 1$ for each~$s
\in \P_1$. 
This determines an interaction exchangeable network, which
we call the \emph{hierarchical vertex components model} (HVCM).
Given $f$, $\I(1),\I(2),\ldots$ are independent, identically
distributed (i.i.d.)~random structured interactions drawn from
\eqref{eq:paintbox}. 
The associated random interaction exchangeable
network~$\Ybf := \ybf_{\I}$ is obtained through \eqref{eq:cong}, whose
distribution we denote by $\epsilon_f$.

In non-HVCMs~\cite{CraneDempsey2016E2}, each constituent element had a
single frequency of occurrence.  
By contrast, HVCMs allow the frequency of occurrence for elements in
the second term of \eqref{eq:paintbox} (i.e., $r \in \P_2$) to depend on first component
(i.e., $\bar s \in \fin \P_1$) through $f_{r|s}$.  This dependence is two-fold: (1) $w
\in \mathcal{F}_1$ controls which $f^{\prime \prime}_{s}$ is chosen
across $s \in \bar s$; and (2) the local distributions can vary,
leading to the size-biased ordering of $r \in \P_2$ varying as a
function of $s$.  

\begin{rmk}[Vertex exchangeability versus interaction exchangeability] \normalfont
While HVCMs are expressed as a function of the vertices, they are
interaction exchangeable and not vertex exchangeable.  To see
this, consider Theorem~\ref{thm:repthm}.  A direct corollary is that
vertices are sampled in size-biased order according to their relative
frequency of occurrence. In hierarchical models, the size-biased
sampling of the second component depends on the first component. Regardless,
this implies the observed constituent elements are not exchangeable
with the unobserved constituent elements.  
On the other hand, vertex exchangeability implicitly assumes the observed vertices
and unobserved vertices are exchangeable.
\end{rmk}

\section{Sequential description for particular subfamily of HVCMs} \label{section:emailmodel}
% \textcolor{red}{So should we change names? Al will continue to suggest we do so,
% maybe I can bug him for suggestions.}
Here we provide a sequential description of a particular subfamily of
HVCMs.
For ease of comprehension, we start with the setting of a single sender where the size of
the first component is one (i.e., $\nu_{k_1} = 1[k_1 = 1]$).
In this setting, the sequential description is presented in the
context of e-mails. 
Let~$(\tilde\alpha, \tilde\theta)$ satisfy either (1) $0 \leq \tilde
\alpha < 1$ and $\tilde\theta > 0$, or (2) $\tilde \alpha<0$ and
$\theta = - K \tilde\alpha$ for some $K\in\Nat$.  In setting~(1), the 
population~$\P_1$ is infinite, while in setting~(2) the population is
finite and equal to~$K$. 
In Section~\ref{section:articlemodel}, we show how to extend this
model to the general setting of multiple senders. For ease of comprehension, we let~$\S =
\P_1$ (senders) and~$\R = \P_2$ (receivers) denote the two sets of
constituent elements. 

We introduce some additional notation.
For each~$n=1,2,\ldots$, the $n$th email~$E_n$ is given by the 
structured interaction~$(\bar S_n, \bar R_n ) = ( \{ S_{n,1} \}, \{
R_{n,1}, \ldots, R_{n, k_{n,2}} \} )$ where~$S_{n,1} \in \S$ is the
sender, and~$R_{n,j} \in \R$ is the~$j$th receiver of the $n$th
article.
% and~$k_n$ denotes the number of recipients.
% \textcolor{red}{I'm going to change sender degree to Dout, to more reflect
%   network sciency lit. Will probably make stats. networks. people
%   happy to see something familiar.}
Suppose $n$ articles have been observed and define $\H_{n} = \left \{
  E_1, \ldots, E_{n} \right \}$ to be the observed history of the
first $n$ e-mails.  For the~$(n+1)$st e-mail, choose the sender
according to 
\begin{equation}
\label{eq:hollywood}
  \pr \left( S_{n+1,1} = s \given \H_{n} \right)
  \propto  \left \{
    \begin{array}{c c} 
      D^{out}_{n} (s) - \tilde\alpha
      & s \in \S_{n} \\
      \tilde\theta + \tilde\alpha | \S_{n} |
      & s \not \in \S_{n}.
    \end{array}
  \right.
\end{equation}
where~$D_n^{out}(s)$ is the outdegree of the subject~$s$,
and~$\S_{n}$ are the set of unique senders in~$(\bar S_1,\ldots,
\bar S_n )$ and~$|\S_n|$ is the set's cardinality. 
% We write~$\S_{n}$
% to denote the unique senders in~~$(\bar_1,\ldots, \bar S_n )$.

% Next define a random variable~$Z_n$ with domain $\bar S_n$.
% Define~$\S^{(z)}_{n}$ to be the unique elements in $\H^{(z)}_n :=
% (Z_1, \ldots, Z_{n})$.  Then
% \begin{equation}
% \label{eq:latenthollywood}
%   \pr \left( Z_{n} = s \given \H^{(z)}_{n}, \bar S_n \right)
%   \propto  {\bf 1} \left \{
%     \begin{array}{c c} 
%       D^{(z)}_{n} (s) - \tilde\alpha_z
%       & s \in \S_{n}^{(z)} \cap \bar S_n \\
%       \tilde\theta_z + \tilde\alpha_z | \S^{(z)}_{n} |
%       & s \not \in \S^{(z)}_{n} \cap \bar S_n \\ 
%       0 & s \not \in \bar S_n 
%     \end{array}
%   \right.
% \end{equation}
% If~$Z_{n} = s$ for $s \in \S^{(z)}_{n}$, then increase~$D_{n}^{(z)}
% (s)$ by one. If~$s \not \in \S_{n}$, then set $D_{n}^{(z)} (s) = 1$.

Given~$S_{n+1,1} = s \in \mathcal{S}$, we choose the number of
recipients according to the discrete probability distribution 
function~$\{ \nu_{k}^{(s)} \}_{k=1}^{\infty}$.
% We now provide additional notation required for describing the
% probability distribution for recipient selection.
% Define~$R_{n,j} (s) \subset \mathcal{R}$ 
% be the set of receivers for e-mails by 
% sender~$s$ up to but not including the $j$th recipient of  
% the~$n$th e-mail; that is,
% \[
% R_{n,j} (s) = \{ r \in \mathcal{R} \, | \, \exists \, R_{m,l} =  r
% \text{ and } S_m = s \text{
% for } m < n \, \& \, l \leq k_m \text{ or } m = n \, \& \, l < j \}
% \]
Finally, let~$D_{n,j} (s,r)$ denote the indegree of receiver~$r$
when restricted to e-mails from sender~$s$ after the first~$n$
e-mails and the $j-1$ recipients of the $n$th e-mail; that is, 
\[
  D_{n,j} (s,r) = \# \, \{ (m,l) \, | \, R_{m,l} =  r
  \text{ and } S_{m,1} = s \text{
    for } m < n \; \text{and} \; l \leq k_m \text{, or } m = n \; \text{and} \; l < j \}.
\]
Finally, we define~$m_{n,j} (s) = \sum_{r \in \mathcal{R} } D_{n,j}
(s,r)$ to be the number of receivers (accounting for multiplicity)
of e-mails from sender~$s$. Each of these statistics 
% given the e-mail history
% up to the~$j$th recipient of the $n$th e-mail
% \[
% \H_{n,j} = \left \{ E_1, \ldots, E_{n}, \left ( \bar S_{n+1} , \{
%     R_{n+1,1}, \ldots, R_{n+1,j} \} \right) \right \}. 
% \] 
% i.e., $D_{n,j} (r,s)$ and~$m_{n,j} (s)$, 
is a measurable function of~$\H_{n}$.  Note, these statistics are
\emph{local} (i.e., specific to the particular subject).  Here, we
describe a procedure for sharing information across senders. To do
this, we define a \emph{partially observable} global set of
information. First, define the observable variable~$R_{n,j}$ to be the
complete set of receivers; that is,
\[
R_{n,j} = \{ r \in \mathcal{R} \, | \, \exists \, R_{m,l} =  r
\text{ for } m < n \, \text{and} \, l \leq k_m \text{, or } m = n \, \text{and} \, l < j \}.
\]
Additionally, let $K_{n,j} = |R_{n,j}|$ be the cardinality of this
set.  For each~$r \in R_{n,j}$ we posit existence of a \emph{latent
  degree} per sender~$s \in \S_n$ and receiver $r$ denoted by~$V_{n,j}
(s,r)$. We then define~$V_{n,j} (\cdot, r) = \sum_{s \in \S_n} V_{n,j} (s, r)$
and $m_{n,j} = \sum_{r \in \R_{n,j}} V_{n,j} (\cdot, r)$.
Next, define~$R_{n,j}(s)$ to be the complete set of receivers when
restricting to e-mails from sender~$s \in \S_n$, and~$\H_{n,j}$ to be
the observable history~$\mathcal{H}_{n-1}$ union $\{ S_{n,1}, R_{n,1},
\ldots, R_{n,j} \}$.  That is,~$\H_{n,j}$ is the observed history up
to the $j-1$th receiver on the $n$th e-mail, where~$\H_{n,0}$ implies
only sender information for the $n$th e-mail.
Finally, for each~$s \in \S$, let~$(\alpha_s, \theta_s)$ satisfy
either (1) $0 \leq \alpha_s < 1$ and $\theta > 0$, or (2) $\alpha_s<0$ and
$\theta_s = - K^\prime \alpha_s$ for some $K^\prime \in\Nat$.  In setting~(1), the 
receiver population~$\P_2$ is infinite, while in setting~(2) the
population is finite and equal to~$K^\prime$. For the remainder of
this paper, we assume setting (1).

Given the indegree distribution~$\{ D_{n,j} (s^\prime,r^\prime) \}_{r^\prime
  \in R_{n,j}, s^\prime \in \S_n}$, the latent degree distribution
~$\{ V_{n,j} (s^\prime, r^\prime) \}_{r^\prime \in R_{n,j}, s^\prime
  \in \S_n}$, the current sender $s$, along with the observable
history~$\H_{n,j}$, the probability of choosing receiver~$r$ is
proportional to
\begin{equation}
\label{eq:seqenron1}
  % \pr (R_{n,j} = r \given H_{n,j}, \{ D_{n,j} (r) \}_{r \in R_{n,j}}, 
  % \{ V_{n,j} (r, \cdot) \}_{r \in R_{n,j}} ) \propto 
  % \left \{
  %   \begin{array}{c c}
\frac{D_{n,j} (s,r) - \alpha_s V_{n,j} (s,r) + (\theta_s +
  \alpha_s V_{n,j} (s,r) ) \left( \frac{V_{n,j} (\cdot, r) -
      \alpha}{m_{n,j} + \theta} \right)} {m_{n,j}(s) + \theta_s}, 
r \in R_{n,j} (s) 
\end{equation}
and
\begin{equation}
\label{eq:seqenron2}
\frac{\theta_s + \alpha_s V_{n,j} (s,r) }
{m_{n,j} (s) + \theta_s} \cdot 
\frac{\theta  + \alpha V_{n,j}(\cdot, r) }{m_{n,j} + \theta}, r \not \in
R_{n,j} (s).
\end{equation}
  %   \end{array}
  % \right.
Note the difference in the discount of indegree
in~\eqref{eq:seqenron1} and outdegree in~\eqref{eq:hollywood}.
For the sender distribution~\eqref{eq:hollywood}, the outdegree
discount is~$\tilde \alpha$; on the other hand,
for~\eqref{eq:seqenron1}, the indegree discount is~$\alpha_s
V_{n,j}(s,r)$.  This reflects that in~\eqref{eq:hollywood}, sender~$s$
is chosen from a single distribution; however,
in~\eqref{eq:seqenron1}, receiver~$r$ can be chosen either locally or
globally. 
% Discounting by $V_{n,j} (r,s)$ accounts for these dual
% possibilities.

The remaining question is how to update the degree
distributions. In~\eqref{eq:seqenron1} and~\eqref{eq:seqenron2}, we
can either observe~$r$ ``locally'', or we escape the local model and
observe~$r$ due to the latent global information. Given~$R_{n,j} = r$
we update both local and global degrees. If~$r \not \in R_{n,j} (s)$
then the global degree~$V_{n,j} (s,r)$ increases from zero to
one. If~$r \in R_{n,j} (s)$ then the local degree~$D_{n,j} (s,r)$
increases by one and the latent degree is increased by one with
probability $\tau_{n,j} (s) = \frac{\theta_s + \alpha_s V_{n,j}
  (s,r)}{m_{n,j} (s) + \theta_s}$. The exact procedure for incrementing
$V_{n,j}$ is discussed in section~\ref{section:estimation}.
% , and this exact
% formulation is called the \emph{extended Enron model}.
% If global degree increases, increase~$V_{n,j} (r,s)$ by one.  

\subsection{Partial pooling}

The importance of the latent global degree distribution is that it allows
information to be shared across the conditional receiver
distributions. The above model formalizes the partial pooling of
information. The degree of pooling is controlled by the escape
probability~$\tau_{n,j} (s)$, which in general decreases as the number
of e-mails from sender~$s$ increases. Note that over time as more
e-mails by sender~$s$ are seen, the escape probability~$\tau_{n,j} (s)$
tends to zero whenever $\alpha_s < 1$.
Therefore, the local impact of the latent global degree information
becomes negligible once we have sufficient local information.
% for pooling to be unnecessary.
However, the first time a sender-receiver pair is observed, it 
must occur via the shared global set of information.
The global latent degrees~$\{ V_{n,j} (s,r) \}_{r \in R_{n,j}, s \in S_n}$ therefore 
contribute to the behavior of new and/or rarely seen senders.  

%\subsection{Interpretation of parameters and finite population model}

% \begin{itemize}
% \item $\alpha$, $\theta$ are the global parameters.  
% \begin{itemize}
% \item $0 < \alpha < 1, \theta > 0$: infinite set of receivers.  $\alpha$ controls
%   power-law (Thm 1).  Whereas $\theta$: controls rate of convergence to power law
% \item $\alpha = 0, \theta > 0$: infinite set of receivers, no power law
% \item $\alpha < 0, \theta = -K \alpha$: finite set of $K$ receivers.
% \end{itemize}
% \item For each~$s \in \S$, $\alpha_s$ and $\theta_s$ control two
%   things
% \begin{itemize}
% \item Exit probabilities -> 0 regardless.  But faster for $\alpha_s$
%   near one and $\theta_s$ near zero.
% \item Local behaviors...
% \item We can have infinite set of receivers but each local model can
%   be finite!
% \end{itemize}
% \end{itemize}

\subsection{Connection between sequential description and hierarchical
  vertex components models}\label{section:vert-comp-ESF}

The sequential description in section~\ref{section:estimation} is
equivalent to a particular HVCM. When~$\alpha_s = 0,~ \forall s \in \P_1$, an analytic
stick-breaking representation can be derived. This connects the
sequential process directly to~\eqref{eq:paintbox}. To do so, we start
by constructing the sender distribution. Here, we assume~$\P_1 \equiv
\P_2 \equiv \Nb$. For~$s \in \mathbb{N}$, define independent random
variables~$\beta_s \sim \text{Beta} (1-\tilde{\alpha}, \tilde{\theta}
+ s \tilde{\alpha} )$. Then, conditional on~$\{\beta_s\}_{s=1}^{\infty}$,
the probability of choosing sender~$s \in \mathbb{N}$ is given by
\[
f_s \given \{\beta_{s^\prime}\}_{s^\prime=1}^{\infty} = \beta_s \prod_{i=1}^{s-1} (1-\beta_i),
\]
where the product is set equal to one for~$s = 1$, and $f^\prime =
\{f_s\}_{s=1}^\infty$. 
In our current setting, $\nu_{k_1} = 1[k_1 = 1]$ so the weights~$w =
\{ w_s \}_{s=1}^\infty$ can be ignored for now.  See Section~\ref{section:articlemodel}
for a description of how these can be constructed in a similar manner. 

We now construct, for each $s \in \mathbb{N}$ the probabilities~$\{ f_{r \given s}
\}_{r=1}^\infty$ via a hierarchical model given $\alpha > 0$ and
$\theta > -\alpha$, and we set $f^{\prime \prime} = \{\{f_{r|s}\}_{r=1}^\infty\}_{s=1}^\infty$ .
To do this, we first define global independent random
variables~$\tilde \beta_r \sim \text{Beta} ( 1-\alpha, \theta + r
\alpha )$ for $r\in \mathbb{N}$. 
Then, conditional on~$\{\tilde \beta_r \}_{r=1}^\infty$,
for $r\in\mathbb{N}$, we define associated stick-breaking
probabilities~$\tilde{\pi}_r = \tilde \beta_r \prod_{i=1}^{r-1} (1 -
\tilde \beta_{i})$.
These are probabilities of choosing receiver~$r$ based on the 
global random variables~$\{\tilde \beta\}_{r=1}^\infty$.
The local stick-breaking distributions are then defined via a
perturbation of these global probabilities. That is, for~$\theta_s >
0$, define independent random variable
\begin{align*}
\tilde \beta^\prime_{r \given s} &\sim \text{Beta} \left( \theta_s
  \tilde \pi_{r}, \theta_s \left( 1 - \sum_{l=1}^{r} \tilde \pi_l
  \right) \right) \\
f_{r \given s} \given \{ \tilde \beta^\prime_{j \given s}
  \}_{j=1}^\infty &= \tilde \beta^\prime_{r \given s}
                    \prod_{i=1}^{r-1} \left( 1 - \tilde \beta_{i
                    \given s}^\prime \right)
\end{align*}
where the product is defined equal to one when~$r = 1$.
This yields a stick breaking representation for $f = \{f^\prime, f^{\prime
  \prime}\}$ for a particular
hierarchical vertex components model. Partial pooling occurs via the
shared global probabilities~$\tilde \pi_{r}$.  The local distributions
satisfy $\EE \left [ f_{r \given s} \given \{ \tilde{\pi}_{r^\prime}
  \}_{r^\prime=1}^{\infty} \right] = \pi_r$.  Therefore, the
distribution $\{ f_{r \given s} \}_{r=1}^{\infty}$ can be thought of
as perturbation of the global distribution~$\{ \pi_{r}
\}_{r=1}^{\infty}$ where~$\theta_s$ controls the amount of
perturbation. In particular, $f_{r \given s} \to \pi_r$ with
probability one as~$\theta_s \to \infty$.

By construction,~$f = (\{ f_s \}_{s=1}^\infty, \{ f_{r \given
  s}\}_{s,r=1}^\infty)$ is a random variable over the space
$\mathcal{F}_1 \times \otimes_{s=1}^{\infty} \mathcal{F}_2$.
Lemma~\ref{lemma:stickbreaking} establishes the connection between
these random variables and the canonical model for $\alpha_s = 0$.  
Although this model for $\alpha_s > 0$ does not admit a known stick
breaking representation, Theorem~\ref{thm:repthm} discussed in Section
\ref{section:markove2} ensures these asymptotic frequencies
exist. Section~\ref{appendix:technical} of the supplementary materials
describes a specific probabilistic construction of these frequencies.  

\begin{lemma} \label{lemma:stickbreaking}
  The sequential HVCM model for~$\alpha_s = 0$ for $s \in \mathbb{N}$ is equivalent in
  distribution to (3), where $f$ is distributed according to the stick-breaking
  construction described above.
\end{lemma}

Proof can be found in section~\ref{appendix:technical} of the
supplementary materials; see~\cite{Pitman2005,Ishwaran2001} for
further details on the stick-breaking representation. 

\begin{rmk}[Connections to CRFP] 
\normalfont
  Note that the HVCM described above is closely related to the
  Chinese Restaurant Franchise Process, a well-known process in the
  machine learning literature~\cite{Pitman2005, blei2003latent} that
  is almost exclusively used to model
  latent clusters in data. Here, we use these ideas in the
  construction of the interaction process. Thus, the
  objectives are quite different; for instance, there is almost no
  focus on the inference of the model parameters in the ML community;
  in our setting, these parameters are crucial to understanding the
  overall interaction process behavior. This model is most similar
  to~\cite{Teh06abayesian}, where it is used for language modeling. Similar to
  the CFRP, the above construction is related to the Pitman-Yor process and the
  GEM distributions~\cite{Pitman2005}. More details can be found in
  Section\ref{section:estimation} and the supplementary materials.
\end{rmk}

\subsection{Accounting for multiple elements in first component}
\label{section:articlemodel}
In the general setting, the first component,~$\bar S_n$, is a random
element of $\fin (\P_1)$ (i.e., a random finite multiset of elements
from $\P_1$).  In the sequential description, we assumed the size of
this multiset was one.  We now consider~$\bar S_n = \{ S_{n,1},
\ldots, S_{n,k_{n,1}} \}$ for general $k_{n,1} \geq 1$.  
First, let~$\H^{(s)}_{n,j} = \H_{n} \cup \{S_{n+1,1}, \ldots, S_{n+1,j} \}$
denote the history of the first $n$ e-mails and the first $j$ senders
of the $n+1$st e-mail. 
Extension of~\ref{eq:hollywood} to handle multiple senders is
straightforward by replacing $\H_{n}$ by $\H_{n,j}^{(s)}$ and defining
all other terms similarly.

In the sequential description, the sender $S_{n,1}$ is used to specify
which local statistics (i.e., $V_{n,j} (r,s)$, $D_{n,j} (r,s)$ and
$m_{n,j} (s)$) to consider.  However, when there are multiple senders,
this choice is no longer straightforward. To address this, we
introduce a random variable~$Z_n$ with domain $\bar S_n$. This
variable indicates which local statistics will be used in receiver
distributions (i.e., equations~\eqref{eq:seqenron1}
and~\eqref{eq:seqenron2}).  Define~$\S^{(z)}_{n}$ to be the unique
elements in $\H^{(z)}_n := (Z_1, \ldots, Z_{n})$.  Then
\begin{equation}
\label{eq:latenthollywood}
  \pr \left( Z_{n} = s \given \H^{(z)}_{n}, \bar S_n \right)
  \propto  {\bf 1} \left \{
    \begin{array}{c c} 
      D^{(z)}_{n} (s) - \tilde\alpha_z
      & s \in \S_{n}^{(z)} \cap \bar S_n \\
      \tilde\theta_z + \tilde\alpha_z | \S^{(z)}_{n} |
      & s \not \in \S^{(z)}_{n} \cap \bar S_n \\ 
      0 & s \not \in \bar S_n 
    \end{array}
  \right.
\end{equation}
where (1)~$0 < \tilde\alpha_z < 1$ and $\tilde\theta_z > 0$ if the 
population~$\S$ is considered infinite, and (2)~$\tilde \alpha_s < 0$ 
and $\tilde \theta_z = - K \tilde \alpha_z$ if population is finite
and $|\S| = K$.
This is equivalent to restricting~\eqref{eq:hollywood} to be non-zero
only on the domain $\bar S_n$.
Moreover, it is conditional on the history $\H^{(z)}_n$ instead of
$\H_n$. If~$Z_{n} = s$ for $s \in \S^{(z)}_{n}$, then
increase~$D_{n}^{(z)} (s)$ by one. If~$s \not \in \S_{n}$, then set
$D_{n}^{(z)} (s) = 1$.

\section{Statistical properties} \label{section:properties}

We now state several theoretical results for the proposed HVCM
built from the sequential description in
section~\ref{section:emailmodel}. For ease of comprehension, we refer
to this model as the ``canonical HVCM model''.

\begin{thm}\label{thm:e2}
The canonical HVCM with parameters~$\Psi = (\tilde{\alpha},
\tilde{\theta}, \alpha, \theta, \{ \alpha_s, \theta_s \}_{s \in \P_1})$
determines a structured exchangeable interaction  probability
distribution for all~$\Psi$ in the parameter space.
\end{thm}

\noindent Theorem~\ref{thm:e2} is not immediate from the
sequential construction in section~\ref{section:emailmodel}, but is
clear from the reparameterization of the model presented in
section~\ref{section:estimation}, and its connection to the model
previously discussed (this is formalized in
section~\ref{appendix:technical}) of the supplementary materials. 

The remainder of this section focuses on the setting where the size of
the first component is one (i.e., $\nu_{k_1} = 1[k_1 = 1]$). Moreover, we will make
certain alternative assumptions concerning the sender distributions.
These constraints allow sufficient complexity to be interesting, but
assume sufficient regularity to push through the theoretical analysis. First, we
turn to the growth rates in the expected number of unique
receivers. Unlike the Hollywood model, this rate depends on both the
distribution over senders, the global parameter~$\alpha$, and the
local parameters~$\{ \alpha_s \}_{s\in\P_1}$. Before stating the
theorem, we require a formal definition of sparsity. For clarity, we
define quantities in terms of receivers to distinguish vertices
observed as senders and those observed as receivers (i.e., in $\P_1$
and~$\P_2$ respectively). 

For a structured interaction-labeled network~$\Ybf$, let~$v(\Ybf)$ denote the
number of non-isolated receivers; $e(\Ybf)$ is the number of
interactions;~$M_k (\Ybf)$ is the number of interactions with $k$
receivers; $N_k (\Ybf)$ is the number of receivers that appear exactly
$k$ times; and~$d(\Ybf) = (d_k (\Ybf))_{k \geq 1}$ is the indegree
distribution, where $d_k (\Ybf) = N_k (\Ybf) / v(\Ybf)$.
Note that these are global statistics that do not depend on the
interaction labels.
% These are global statistics (i.e., across senders). 
We define local versions by superscripting each statistic by~$s \in
\P_1$.
For instance, $v^{(s)} (\Ybf)$ is the number of non-isolated receivers
when restricting~$\Ybf$ to only those interactions involving sender~$s$.
The statistics~$e^{(s)} (\Ybf), M^{(s)}_k (\Ybf), N^{(s)}_k (\Ybf),
d^{(s)}(\Ybf)$ and $d_k^{(s)} (\Ybf)$ are defined similarly.

\begin{defn}[Global and local sparsity]
Let~$(\Ybf_{n})_{n \geq 1}$ be a sequence of interaction-labeled networks
for which $e(\Ybf_{n}) \to \infty$ as $n \to \infty$. The sequence
$(\Ybf_{n})_{n \geq 1}$ is \emph{sparse} if
\[
\limsup_{n \to \infty} \frac{e (\Ybf_{n})}{v(\Ybf_n)^{m_{\bullet} (\Ybf_n)}} = 0,
\]
where~$m_{\bullet} (\Ybf_n) = e(\Ybf_n)^{-1} \sum_{k \geq 1} k M_k (\Ybf_n)$ is
the average arity (i.e., number of receivers) of the interactions in $\E_n$.  A non-sparse
network is
~\emph{dense}.
We say the sequence is~$(\E_n)_{n \geq 1}$ is \emph{$s$-locally
  sparse} if
\[
\limsup_{n \to \infty} \frac{e^{(s)} (\Ybf_n)}{v^{(s)} (\Ybf_n)^{m^{(s)}_{\bullet} (\Ybf_n)}} = 0,
\]
where~$m^{(s)}_{\bullet} (\Ybf_n) = e^{(s)}(\Ybf_n)^{-1} \sum_{k \geq 1} k M^{(s)}_k (\Ybf_n)$ is
the average arity (i.e., number of receivers) of the interactions in
$\Ybf_n$ from sender~$s \in \P_1$. A network that is not $s$-locally
sparse is \emph{$s$-locally dense}.
\end{defn}

For~$(X_n)_{n \geq 1}$ a sequence of positive random variables
and~$(y_n)_{n\geq1}$ a sequence of positive non-random variables,
let~$X_n \simeq y_n$ indicate $\lim_{n \to \infty} X_n / y_n$ exists
almost surely and equals a finite and positive random variable.
Theorem~\ref{thm:sparsity} shows the canonical model may be either 
globally sparse and/or dense. The theorem assumes a finite population of
senders with number of e-mails per sender drawn from a multinomial
distribution. 
%This approximates the behavior of the canonical model
% well where  the number of senders is ``large'' (i.e., large relative
% to underlying parameters).
%This mimics stick-breaking representations in clustering, where a
%maximum number of clusters is assumed and a cutoff is applied to the
%prior at the chosen maximum; see~\cite{FoxICMLpaper} for details.

\begin{thm}\label{thm:sparsity}
Suppose the sender population~$\P_1$ is finite, consisting of $d$
senders, i.e., $\P_1 = [d] := \{1,\ldots, d\}$. 
Assume, out of the~$n$ e-mails, the number of e-mails per sender~$s$,
denoted~$n_s$, is drawn from a multinomial distribution with
probabilities~$(p_1, \ldots, p_d)$ such that~$\sum_{s=1}^d p_s = 1$
and~$p_s > 0$ for all $s \in [d]$. 
Let $\mu_s$ be the average size of emails for sender $s$ and
$\mu:=\sum_{s=1}^d p_s \mu_s$ the average size of emails across all
senders. Then 
$v(\Ybf_n) \simeq (\mu^{1/\alpha_\star} \mu_\star p_\star n)^{\alpha_0 \alpha_*}$
% \quad as \, n \rightarrow \infty
%\]
where~$s^\star = \arg \max_{s \in [d]} \alpha_s$,~$\mu_\star =
\mu_{s^\star}$, and~$\alpha_\star = \alpha_{s^\star}$. In particular, 
if~$\mu^{-1} < \alpha \cdot \alpha_\star < 1$, then $(\Ybf_n)_{n \geq
  1}$ is almost surely sparse.
% Moreover, for~$s \in \S$, we have
% \[
% \EE [ v^{(s)} (\Y_n) ] \sim (\mu n)^{\alpha \cdot \alpha_s}
% \]
% and~$\mu = \sum_{k \geq 1} k \nu_k$ is the average arity. Furthermore,~$\mu^{-1} <
% \alpha \cdot \alpha_s < 1$ then $(\Y_n)_{n \geq 1}$ is almost surely
% $s$-locally sparse.
\end{thm}
Theorem~\ref{thm:sparsity} establishes that the canonical HVCM for a special
case of the sender distribution can capture degrees of
sparsity.  If~$\mu_s = \mu$ for all~$s \in \P_1$ and~$\alpha
\alpha_{\star} < \mu^{-1}$ then it must be the case that~$\alpha
\alpha_s < \mu^{-1}$ for all~$s \in \P_1$.  Therefore, a dense network
must be $s$-locally dense for all~$s \in \P_1$.  However, 
a sparse network can be $s$-locally dense for some, but not all, $s
\in \S$. We turn now to considerations of power-law degree
distribution for interaction-labeled networks. We start with a
definition.

\begin{defn}[Global power-law degree distributions] 
A sequence $(\Ybf_n)_{n \geq 1}$ exhibits \emph{power-law degree
  distribution}~\cite{CraneJASA, CaronFox2017, VeitchRoy2015} if for
some~$\gamma > 1$ the degree distributions $(d(\Ybf_n))_{n\geq 1}$
satisfy $d_k(\Ybf_n) \sim l(k) k^{\gamma}$ as $n \to \infty$ for all
large $k$ for some slowly varying function $l(x)$; that is, $\lim_{x
  \to \infty} l(tx)/l(x) = 1$ for all $t > 0$, where $a_n \sim b_n$
indicates that $a_n/b_n \to 1$ as $n \to \infty$.  More precisely,
$(\Ybf_n)_{n \geq 1}$ has power law degree distribution with index
$\gamma$ if
\begin{equation} \label{eq:powerlaw}
\lim_{k \to \infty} \lim_{n \to \infty} \frac{d_k (\Ybf_n)}{l(k)
  k^{-\gamma}} = 1.
\end{equation}
% The sequence $(\E_n)_{n \geq 1}$ has $s$-local power law degree
% distribution with index~$\gamma$ if equation~\eqref{eq:powerlaw}
% holds with~$d_k^{(s)} (\E_n)$ replacing~$d_k (\E_n)$.
\end{defn}

Theorem~\ref{prop:powerlaw} establishes the power-law degree distribution
for the canonical HVCM for the case of $\alpha_s = 1, \forall s \in \mathcal{S}$.
%For other values, simulations presented in
%\textcolor{red}{Appendix~\ref{app:CSNmethod} suggest the result
%  holds across the entire parameter range.}

\begin{thm}\label{prop:powerlaw}
Let~$( \Ybf_n )_{n \in \mathbb{N}}$ obey the sequential description in 
section~\ref{section:emailmodel} with parameters $(\tilde{\alpha},
\tilde{\theta})$ and let $\alpha_s = 1$ for all $s \in \P_1$. 
For each~$n \geq 1$, let $p_n (k) = N_k (\Y_n) /  v (\E_n)$
for~$k \geq 1$ be the empirical receiver degree distribution
where~$N_k (\E_n)$ is the number of receivers of degree~$k \geq 1$
and~$v (\E_n)$ is the number of unique receivers in~$\E_n$.
Then, for every~$k \geq 1$,
\begin{equation} \label{eq:yulelaw}
p_n (k) \sim \alpha k^{-(\alpha + 1)}/ \Gamma (1-\alpha)
\end{equation}
where~$\Gamma(t) = \int_0^\infty x^{t-1} e^{-x} dx$ is the gamma
function. That is,~$(\Y_n)_{n \geq 1}$ has a power law degree
distribution with exponent $\gamma = 1+\alpha \in (1,2)$.

%\textcolor{red}{Suppose~$\alpha_s = 0$ for all~$s \in \P_1$. 
%Let~$\bar f = \{ f_s \}_{s \geq 1} \in \mathcal{F}_1$ be a distribution
%over senders. Assume $\bar f$ satisfies 
%\[
%\text{ conditions }
%\]
%Then, for every~$k \geq 1$, $p_n (k)$ satisfies
%equation~\eqref{eq:yulelaw}. That is,~$(\Y_n)_{n \geq 1}$ has a power
%law degree distribution with exponent $\gamma = 1+\alpha \in (1,2)$.}
\end{thm}

%iTheorem~\ref{prop:powerlaw} shows that for a particular subfamily of
%the Enron process, the global degree distribution is power law
%\emph{regardless} of the local degree distributions.  
%This is akin to....

\section{Posterior inference}\label{section:estimation}

We now consider performing posterior inference for the canonical HVCM 
given an observed interaction network~$\Ybf_n$.
As in Section~\ref{section:emailmodel}, we start with the setting
where the size of the first component is one (i.e., $\nu_{k_1} = 1[k_1
= 0]$). The parameters $(v^{(s)}_{k_2})_{~k_2 \in \mathbb{N}},$ for all $s \in
\mathcal{S}$ are estimated non-parametrically,
and are not important for the remainder of the paper; therefore, the details are
omitted for these parameters.

We start by reparameterizing the HVCM in a more useful form for
inference, and which gives an explicit structure for updating the
latent degree $V_{n,j}$ - we call this the \emph{extended canonical
  HVCM}, or extended model for short.
In this representation, every ``escape'' from the local distribution
and choice of receiver~$r$ leads to an auxiliary vertex~$v$ being
introduced locally for a sender $s$ - auxiliary vertices are not shared between
senders. The \emph{label} $l_s(v)$ of the auxiliary vertex is~$r$; the
auxiliary vertex accounts for the fact that the global distribution
can select receiver~$r$ multiple times. Finally, the observed reciever is
assigned to the auxiliary vertex, and we write that assignment $\phi_{n,j}=v.$
The number of auxiliary vertices with label~$r$ and sender~$s$ is
equal to the number of times the local distribution for sender $s$ escapes and choose the global set of information (i.e.,~$V_{n,j}
(r,s)$). The sum of the degrees across
auxiliary vertices with label~$r$ and sender $s$ is equal to the
indegree for receiver~$r$ (i.e., $D_{n,j} (r,s)$). Finally, we write $d_{srv}$
to denote the degree of auxiliary vertex $v$ in sender $s$ that also
has label $r$. Note that for $r^\prime \neq l_s(v),~d_{srv}=0$. 

Given $\mathcal{H}_{n}$ and $S_{n+1,1} = s$, the probability that
$R_{n+1,j}$ is assigned to auxiliary vertex $\phi_{n+1,j} = v$ is: 
$$
\pr(\phi_{n+1,j} = v \given \mathcal{H}_{n}, S_{n+1,1} = s ) \propto
\begin{cases}
  d_{s \cdot v} - \alpha_s, & v \le V_{n+1,j}(s, \cdot) \\
  \alpha_s V_{n+1,j}(s, \cdot) + \theta_s, & v = V_{n+1,j}(s, \cdot) + 1,
\end{cases}
$$
  
Further, if $\phi_{n+1,j} = V_{n+1,j}(s,\cdot) + 1$, then we add an
auxiliary vertex $V_{n+1,j}(s,\cdot) + 1$ with its label chosen with
probability:
$$
\pr(l_s(V_{n+1,j}+1) \given \phi_{n+1,j} = V_{n+1,j}(s,\cdot) + 1,
\mathcal{H}_{n,j}, S_{n+1} = s) \propto
\begin{cases}
  V_{n,j}(\cdot, r) - \alpha, & r \in \R_{n+1,j} \\
  \alpha V_{n+1,j}(\cdot, r) + \theta, & r \notin \R_{n+1,j}.
\end{cases}
$$
%The extended Enron model requires additional notation which we now
%introduce.
%Let~$R_{n,j}$ continue to denote the~$j$th receiver of the~$n$th
%e-mail, and let~$v_{n,j}$ denote the auxiliary vertex for the~$j$th
%receiver of the $n$th e-mail.
%The previously introduced term~$V_{n,j}(s,r)$ now also denotes the
%number of auxiliary vertices with label~$r$ introduced for sender~$s$.
%Let~$d_{srk}$ denote the degree of the $k$th auxiliary vertex
%($k = 1,\ldots, V_{n,j} (s,r)$) with label~$r$ introduced for sender~$s$.
%Finally, we replace a subscript by $\bullet$ to denote summing over
%this index.
The likelihood of observing~$\Ybf_N = \{\{S_{n,1}, k_n, \{R_{n,j},
\phi_{n,j}\}_{j=1}^{k_n} \}, \{l_s(\cdot)\}_{s\in \S_N}\}_{n=1}^N$
given the parameters~$\Psi = (\tilde{\alpha}, \tilde{\theta}, \alpha,
\theta, \{ \alpha_s, \theta_s \}_{s \in \P_1})$ is given by
\begin{equation}
  \label{eq:extendedEnron}
  \pr(\Ybf_N) = \pr(\{\{R_{n,j}\}_{j=1}^{k_n}\}_{n=1}^N,
    {l_s(\cdot)}_{s \in \S_N} | \{ S_{n,1} \}_{n=1}^N, \{k_n\}_{n=1}^N)
    \pr(\{S_{n,1} \}_{n=1}^N) \pr(\{k_n\}_{n=1}^N),
  \end{equation}
  where
  \begin{align*}
  \pr &\left(\{\{R_{n,j}\}_{j=1}^{k_n}\}_{n=1}^N, {l_s(\cdot)}_{s \in
      \S_N} \given \{S_{n,1} \}_{n=1}^N, \{k_n\}_{n=1}^N \right) \\ 
 &= \frac{[\theta + \alpha]_{\alpha}^{K_N - 1} }{[\theta +
   1]_1^{m_N - 1}} \prod_{r} [ 1 - \alpha ]_1^{V_N(\cdot, r) -
   1} \prod_{s} \frac{[\theta_s + \alpha_s]_{\alpha_s}^{V_N(s, \cdot)
     - 1} }{[\theta_s + 1]_1^{m_N(s) - 1}} \prod_{v=1}^{V_N(s, \cdot)}
 [ 1 - \alpha_s ]_1^{d_{srv} - 1},
\end{align*}
   and
\begin{align*}
  &\pr(\{S_{n,1} \}_{n=1}^N) = \frac{[\tilde \theta + \tilde
    \alpha]_{\alpha}^{\mathcal{S}_N}}{[\tilde \theta + 1]_1^{N}}
    \prod_{s}[1 - \tilde \alpha]_1^{D_{N}^{out}(s)-1}, \\
  &\pr(\{k_n\}_{n=1}^N) = \prod_{n=1}^N v_k^{(s)},
\end{align*}
where~$[a]_b^c = a(a+b)\dots (a+(c-1) b)$ for $c \in \mathbb{N}$
and~$a,b \in \mathbb{R}_+$. The joint density as written
in~\eqref{eq:extendedEnron} is exchangeable with respect to
re-ordering of the interactions.

Lemma~\ref{lemma:marginalization} proves that the proposed canonical HVCM is
recovered by marginalizing over configurations of auxiliary vertex
labels and assignments, which leaves only the observed degrees
$D_{n,j}$ and latent degrees $V_{n,j}$. The complete likelihood for
the canonical model is given in section~\ref{appendix:technical} of
the supplementary materials, and the likelihood is exchangeable,
proving Theorem~\ref{thm:e2}. Proof of
Lemma~\ref{lemma:marginalization} is also left to
section~\ref{appendix:technical} of the supplementary materials. 
%  for presents the likelihood for
% the Enron model after integrating over the auxiliary variables.

\begin{lemma} %[Equivalence via Marginalization]
\label{lemma:marginalization}
Marginalizing the extended model over configurations of auxiliary vertex
assignments and labels recovers the canonical model.
% Tedious part: showing that the sequential enron process has the same
% likelihood as when you marginalize out the extended model and get the
% generalized stirling numbers. (see Antoniak 74)
\end{lemma}

% \textcolor{red}{In the following, we concentrate on inference for the receiver
%   distributions, as inference for the sender distribution can be performed
%   independently; see\cite{CraneDempsey2016E2}}

\subsection{Choice of priors}

Here, we define two approaches to defining priors for the global
parameters $\theta,\alpha$ and local parameters $\theta_s, \alpha_s, s
\in \mathcal{S}$.

\subsubsection{Conjugate Bayesian Parameters}
The first approach is to set the priors for $\theta$ parameters to a high-variance
Gamma distribution, and the priors for the $\alpha$ parameters to the
Beta distribution. In general, the global $\theta$ will be much larger
than the local parameters, and the appropriate values will depend on
the sparsity of the overall network - for instance, the global
$\theta$ for the arXiv data is an order of magnitude greater than the
global $\theta$ for the Enron dataset. An appropriate prior is a prior
distribution such that the posterior predictive checks on sparsity
match the observed degree of sparsity in the interaction data. See
section~\ref{section:ppc} for details on posterior predictive checks
and model comparison.

For datasets of reasonable size, we have found that the prior for the
global parameters does not significantly affect the resulting posterior density.
In the subsequent examples, the size of the datasets was more than sufficient to
not be strongly affected by the choice of global priors.
For the $\alpha$ parameter, this suggests using $\text{Beta}(1,1)$ distribution,
i.e., the Uniform distribution. With $\theta$, different datasets can have
a difference in posterior means that are 2 or 3 orders of magnitude. Although
the posterior density is mostly unchanged, attempting inference with a 
mismatched $\theta$ prior will require more Gibbs samples before mixing occurs. We have found that $\theta \sim \text{Gamma}(1, 10000)$ is
an appropriate diffuse prior that allows for fast mixing. The lower-level
parameter $\theta_s$ is generally much less than the global $\theta$, so 
$\theta \sim \text{Gamma}(1, 1000)$ is an appropriate prior that allows for
variety in distribution but also has a prior mean that is lower than the global
$\theta$. For the local $\alpha_s$, we again use the Uniform distribution. 

\subsubsection{Priors based on Hollywood model fits}
The second approach, which is used in Section~\ref{section:arxiv} for
the arXiv dataset, is to fit the Hollywood
model~\cite{CraneDempsey2016E2} to each of the local datasets, and
then use a~$\text{Gamma}(\hat\theta / 100, 100)$ prior for the
$\theta$s, where $\hat \theta$ is the estimate of $\theta$ for the
Hollywood model. The priors for the $\alpha$'s are again set to
$\text{Beta}(1,1)$.

\subsection{Gibbs sampling algorithm}

Here we introduce a Gibbs sampling algorithm for sampling from the
posterior distribution of~$\Psi$ given an observed interaction-labeled
network~$\Ybf_n$. To do this, we use auxiliary variable
methods~\cite{Escobar1995} to perform conjugate updates for all parameters.
First, define the binary auxiliary variables~$z_{r,j}$ for $r \in \R,
j = 1, \ldots, v_{\bullet r} - 1$ and~$z_{s,r,k,u}$ for $s \in \S,r \in
\R, v = 1, \ldots, V_N(s,\cdot) - 1, u = 1,\ldots, d_{srv}-1$. Next
define auxiliary variables~$y_i$ for $i = 1, \ldots, v(\Ybf_n)-1$
and~$y_{si}$ for $s \in \S$ and $i = 1, \ldots,d_{s \bullet
  \bullet}-1$. Finally, define auxiliary variables~$x, \{ x_s \}_{s\in
  \S} \in [0,1]$.
We formally derive these updates in section~\ref{app:gibbs} of the
supplementary materials; this algorithm is similar to the one
described in~\cite{Teh06abayesian}, except for the modifications
required for our model.
While sampling each auxiliary vertex for the receivers, we also update
the set of auxiliary vertices~$[V_N(s,r)]$ and their
degrees~$d_{srv}$.
\allowdisplaybreaks
\begin{align}
  x &\sim \text{Beta} ( \theta + 1, m_N - 1) \\
  y_i &\sim \text{Bernoulli} \left( \frac{\theta}{\theta + \alpha \cdot
        i} \right), i = 1,\ldots K_N - 1\\
  z_{rj} &\sim \textrm{Bernoulli}\left( \frac{j - 1}{j - \alpha}
           \right), r \in \mathcal{R}_n, j=1,\ldots,V_N(\cdot, r) - 1 \\
  \theta &\sim \textrm{Gamma}\left( \sum_{i=1}^{K_N - 1} y_i + a, b -
           \log x \right)\\
  \alpha &\sim \textrm{Beta}\left( c + \sum_{i=1}^{K_N - 1} (1 - y_i), d
           + \sum_r \sum_{j=1}^{V_N(\cdot, r) - 1} (1 - z_{r,j}) \right)
  \\
  x_s &\sim \text{Beta} \left( \theta_s + 1, V_N(s,\cdot) - 1
        \right), s \in \S_N \\
  y_{si} &\sim \text{Bernoulli} \left( \frac{\theta}{\theta +
           \alpha \cdot i} \right), 
           s \in \mathcal{S}_N, i = 1,\ldots d_{s \bullet \bullet} - 1
  \\
  z_{srvu} &\sim \textrm{Bernoulli}\left( \frac{j - 1}{j - \alpha}
             \right), s \in \mathcal{S}_N, r \in \mathcal{R}_N,
             v=1,\ldots,V_N(s, \cdot) - 1, u=1,\ldots,d_{srv} - 1\\
  \theta_s &\sim \textrm{Gamma} \left(\sum_{i = 1}^{d_{s \bullet \bullet} -
             1} y_{si} + a_s, b_s - \log x_s \right), s \in \mathcal{S}_N
  \\
  \alpha_s &\sim \textrm{Beta}\left( \phi \alpha + \sum_{i=1}^{d_{s
             \bullet \bullet} - 1}(1 - y_{si}), \phi(1 - \alpha) + \sum_r
             \sum_{v=1}^{V_N(s,\cdot) - 1}\sum_u^{d_{srv} - 1}(1 -
             z_{srvu}) \right), s \in \mathcal{S}_n
\end{align}
There are two important differences between this algorithm and
~\cite{Teh06abayesian}. First, in the case of multiple elements in the first
component, we perform an approximate sampling procedure found in
Section~\ref{subsection:approx_collapse} to find the latent $Z_i$. Second,the
language model in~\cite{Teh06abayesian} has multiple levels of hierarchical
parameters, where we have only two
levels of components.
Convergence can be checked via traceplots and, in our experiments,
occurs within the first hundred or so iterations; see
Figure~\ref{fig:enron_trace_plots} for traceplots in the email network
example.

%\subsection{Approximate marginal evidence method for sampling $Z_i$}
\subsection{Approximate sampling in the case of multiple elements in the first component}
\label{subsection:approx_collapse}
In the case~$\bar S_n$ may contain multiple elements, one can
sample from the posterior
$$
\pr(Z_i=s | \mathcal{H}^{(z)}_n, \bar S_i ) \propto \pr(Z_i=s | \bar S_i) \pr
(\bar R_i = \bar r_i | Z_i=s).
$$
Note that, in general, the joint likelihood  $\pr(\bar R_i = \bar r_i | Z_i=s)$ is
difficult to calculate due to the marginalization over all possible
vertex label configurations for $\bar R_i$. Instead, we propose a
sampling procedure to approximate this quantity, by sequentially sampling the
vertex labels $\bar V_i$ using the given counts, where $\bar V_i$
denotes the multiset $V_{i,1},\ldots, V_{i,k_{i,2}}$:  
$$
\pr (\bar R_i = \{ R_{i,1},\ldots, R_{i,k_{i,2}} \}, \bar V_i |
\mathcal{H}^{(z)}_n, Z_i) = \prod_{j=1}^{k_{i,2}}
\pr(R_{i,j}=r_{i,j}, V_{i,j} = v_{i,j} | \mathcal{H}^{(z)}_n, R_{i,j-1}=r_{i,j-1},
R_{i,j-1}=r_{i,j-1}, \ldots,Z_i=s ).
$$
After sampling $\bar V_i$ for a number of runs, we average the likelihoods to
get an estimate of  $\pr(\bar R_i = \bar r_i | Z_i)$.

\section{Application to Enron email network} \label{section:hierarchical_application}
In this section the proposed HVCM model and inference procedure is applied to the Enron email
dataset. Further, techniques to demonstrate the goodness of fit of the model are
discussed, and are applied in comparison with 
with the previously published ``Hollywood'' model~\cite{CraneJASA} and the generalized gamma process (GGP)
model~\cite{CaronFox2017}; in particular, the HVCM model is shown to have better
model fit at the local level compared to others. %Finally, 
%the section ends with a discussion on the variation that is captured with the
%hierarchical exchangeable model that is not possible with other models that do
%not respect the complete structure of the data.

\subsection{Dataset overview} \label{subsection:enron_overview}
The Enron email dataset consists of approximately 500,000 emails collected from
1998 to 2002 and was originally collected by the Federal Energy Regulation
Commission during its investigation into the company~\cite{Cohen2009}. The
dataset originates from an email dump of 150 users. In total, there are 19,752
unique senders, 70,572 unique receivers, for a total of 79,735 unique entities.
The dataset has been used as a testbed for classification~\cite{klimt2004enron}, topic
modeling~\cite{mccallum2007topic}, and graph-based anomaly
detection~\cite{priebe2005scan, silva2009hypergraph}, among other tasks. 

\begin{figure}[ht]
\centering
\includegraphics[width=2in]{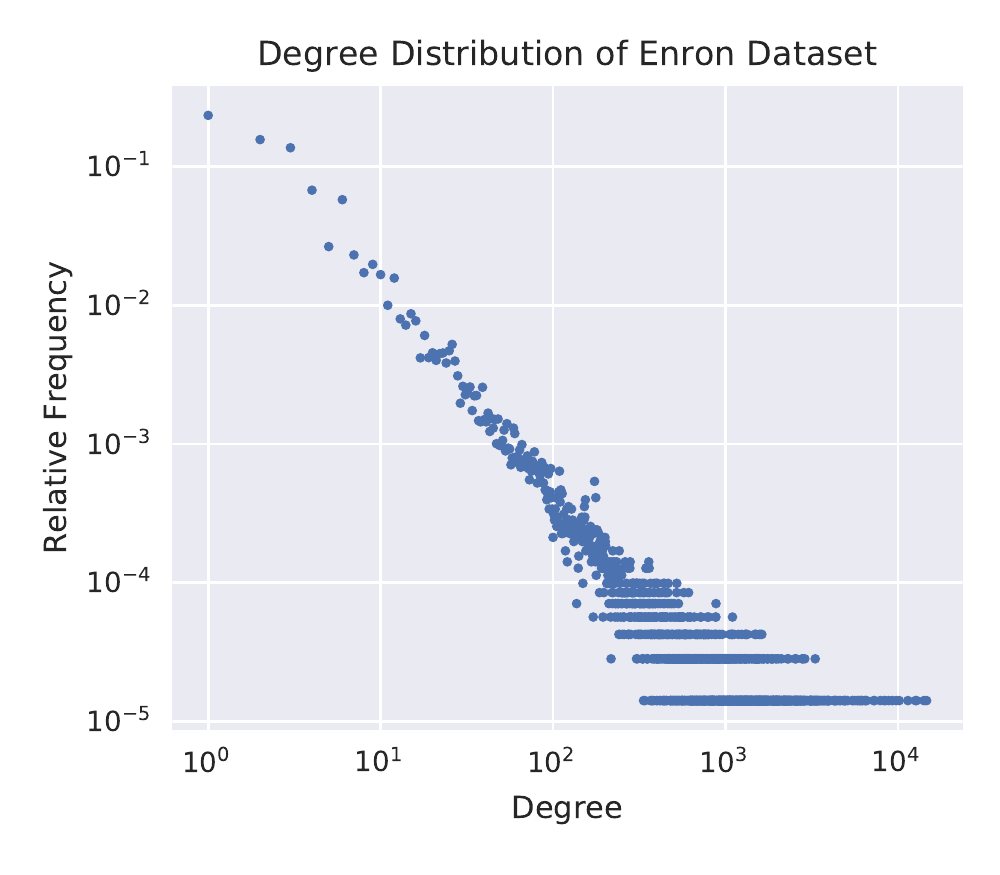}~
\includegraphics[width=4in]{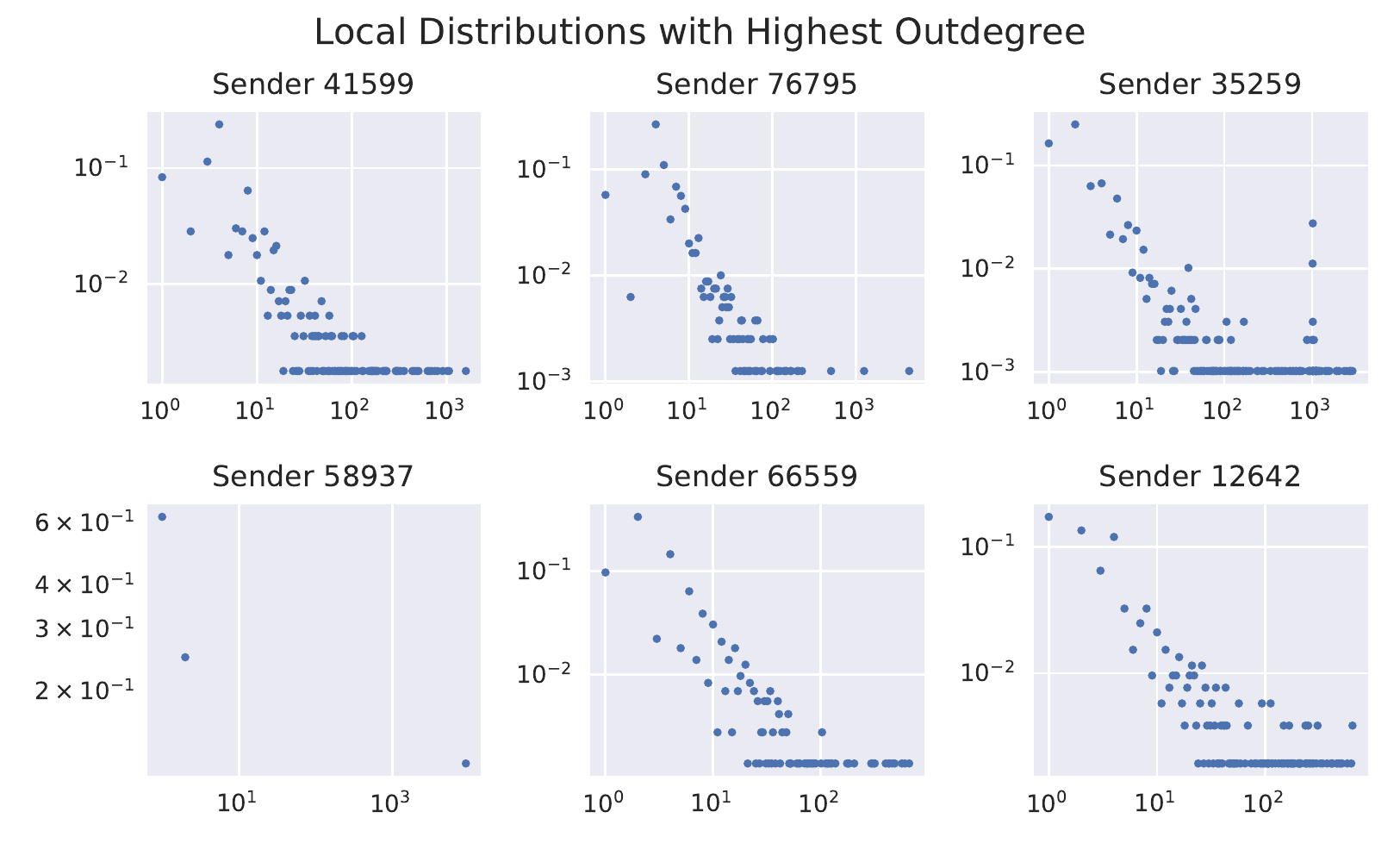}
\caption{Global receiver distribution (left) and some examples of local degree
  distributions (right). There is variation in the shape of these distributions;
  the HVCM accounts for and parameterizes this difference in behavior
  when compared with the global degree distribution.}
\label{fig:enron_deg_dist}
\end{figure}

Figure~\ref{fig:enron_deg_dist} shows the global receiver degree distribution,
as well as the local receiver distributions for the six senders with the largest
number of emails. There is significant variation in behavior of the local degree
distributions, both in comparison to themselves and to the behavior of the
global degree distribution. This suggests that a modeling approach that allows for these
differences is critical to accurately capturing the behavior of the entities,
and thereby allowing for superior data summarization, sound inferences and strong prediction performance. While
the Hollywood and GGP model would be unable to account for this variation, the
proposed HVCM is equipped to capture this behavior.

%These local distributions do not perfectly follow a power-law distribution; a glaring instance of this is Sender 58397, which has very few receivers, and clearly does not follow a power-law. Other outliers are clearly present in these local distributions. This will become more apparent when we compare to synthetic data in order to perform posterior predictive checks in Figure~\ref{fig:enron_compare_local}. 

\subsection{Fit to the data}
%\begin{figure}[ht]
%\centering
%\includegraphics[width=2.5in]{figs/trace_plots_enron.eps}
%\includegraphics[width=2.5in]{figs/hist_alpha_theta_enron.eps}
%\caption{Trace plots and histograms for global parameters of the Enron data. After about 100 iterations, we see tha%t the sampler gives reliable posterior samples.}
%\label{fig:enron_trace_plots} 
%\end{figure}

\begin{figure}[ht]
\centering
\includegraphics[width=4.8in]{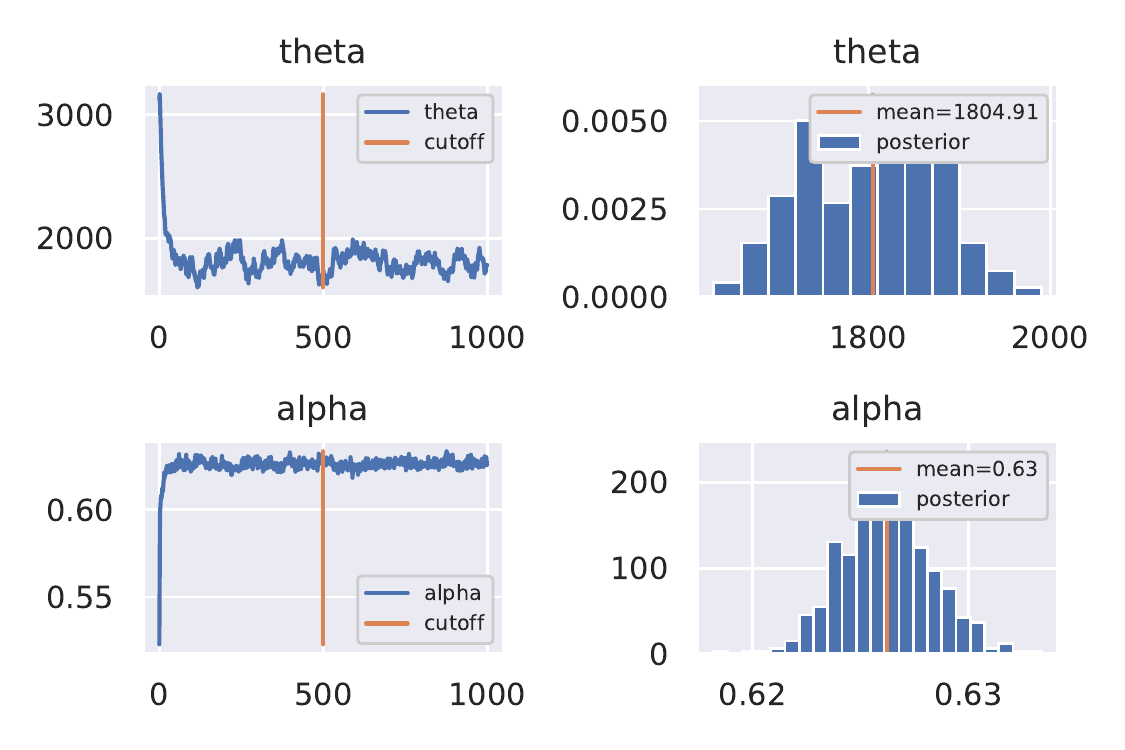}

\caption{Trace plots and histograms for global parameters of the Enron data.
  Mixing occurs after 50 iterations or less. For the posterior predictive
  checks, the last 500 posterior samples were used.}
\label{fig:enron_trace_plots}
\end{figure}

%\begin{figure}[ht]
%\centering
%\includegraphics[width=4in]{img/hist_alpha_theta_enron.eps}
%\caption{Posterior distributions for $\alpha, \theta$.}
%\label{fig:enron_param_hist}
%\end{figure}

The Gibbs sampling algorithm introduced Section~\ref{section:estimation} is applied to the
dataset for 1000 iterations, discarding the first 500 as
burn-in. For this dataset, the following priors were used:
\begin{align*}
  &\pr(\theta) \sim \text{Gamma}(2, 1000),  &\pr(\alpha) \sim \text{Beta}(1,1) \\
  &\pr(\theta_s) \sim \text{Gamma}(1,20),  &\pr(\alpha_s) \sim \text{Beta}(1, 0.9)
\label{eq:enron_priors}
\end{align*}

Trace plots and histograms of the posterior samples of the 
global parameters $\alpha$ and $\theta$ are displayed in
Figure~\ref{fig:enron_trace_plots}. Note that discarding 500 posterior samples
as burn-in is rather conservative, as the Gibbs sampler sampled chain mixes in
less than 100 iterations.

\begin{figure}[ht]
\centering
\includegraphics[width=6.3in]{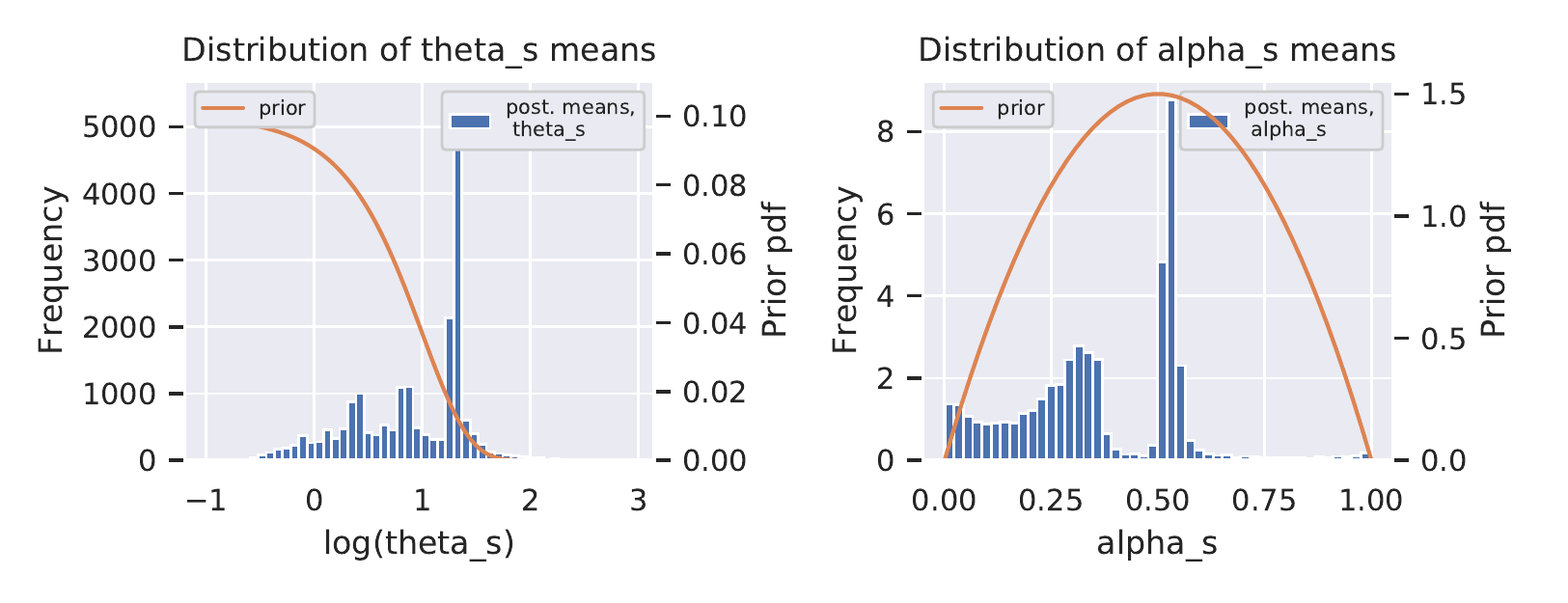}
\caption{Histograms of local $\alpha_s$ and $\theta_s$. Prior pdfs are shown in
  orange. The $\theta_s$ prior is set to fit the local
  distributions; the $\alpha_s$ posterior means are robust to the prior
  distribution chosen.}
\label{fig:enron_local_parameters}
\end{figure}

We show the histogram of posterior means of the local parameters $\theta_s$ and $\alpha_s$ in
Figure~\ref{fig:enron_local_parameters}, along with their priors. The $\theta_s$
parameters are shown on a log scale. These local histograms show significant
diversity among the posterior parameter estimates, as we are fitting local
variations in behavior. For $\alpha_s$, the choice of prior has very little effect on
the posterior samples, except in the case of a small amount of local data for that particular sender
$s$. The choice of prior for $\theta_s$ has more influence on the posterior distribution; our prior of
$\text{Gamma}(1,20)$ is set to bias the local $\theta_s$ towards 0; this will allow for a better fit on the
local data than a prior with larger variance or mean; this result is borne out
when posterior predictive checks are applied to the local sender distributions,
i.e., Figures~\ref{fig:num_verts_enron} and~\ref{fig:enron_compare_local}.

\subsection{Posterior Predictive Checks (PPC) and Model Comparison} \label{section:ppc}
In this section, examples of posterior predictive model checks are shown in order to demonstrate
the goodness of fit of the proposed HVCM. Posterior predictive checks are often used in order to verify that the proposed fitted model generates
reasonable values on statistics of interest; these checks can also be used to
diagnose where the model fails to perform well~\cite{Gelman1996}.

Multiple synthetic datasets are generated according to the posterior predictive distribution as prescribed
in~\cite{Gelman1996}, and statistics of interest are calculated and compared
with the statistics of the real data. The synthetic data is generated from the model with
the parameters set to a posterior sample generated from the inference procedure. Since we
are interested in the ability of the model to account for variation in local behavior, we take the sender
sequence and number of receivers for each email as given, in order to directly compare the
local receiver distributions of the posterior predictive data with the real
data.

In addition to generating posterior predictive checks for the fitted
HVCM, they are also generated for the Hollywood~\cite{CraneJASA} and
GGP~\cite{CaronFox2017} models for comparison. In the following
subsections, a variety of posterior predictive statistics are
described, both for the global dataset and for the local data per sender. These
checks show that the proposed HVCM both provides a good global fit of
the data, in addition to significantly improving the fit to local
distributions compared to the Hollywood
model. Table~\ref{table:ppc_global_enron} details the 95\%
posterior predictive intervals for the global statistics, and
Table~\ref{table:ppc_local_enron} summarizes the posterior
predictive coverage rate for local distributions for the proposed model and the
Hollywood model.
The statistics compared are number of unique receivers in the dataset and number of
receivers with degree 1, 10, and 100.
%documenting the amount of local distributions \textcolor{red}{of a
%  given degree} that are covered by the posterior predictive
%intervals.

%\textcolor{blue}{We need to explain the tables better.  Are these the number of
%  unique senders with degree (of unique receivers) equal to these
%  various numbers)? Are ``Num Vert'' the number of unique receivers in
%the entire dataset?}

\subsubsection{Number of unique receivers}
\label{subsubsection:unique_verts}
\begin{figure}[ht]
  \centering
  \includegraphics[width=5in]{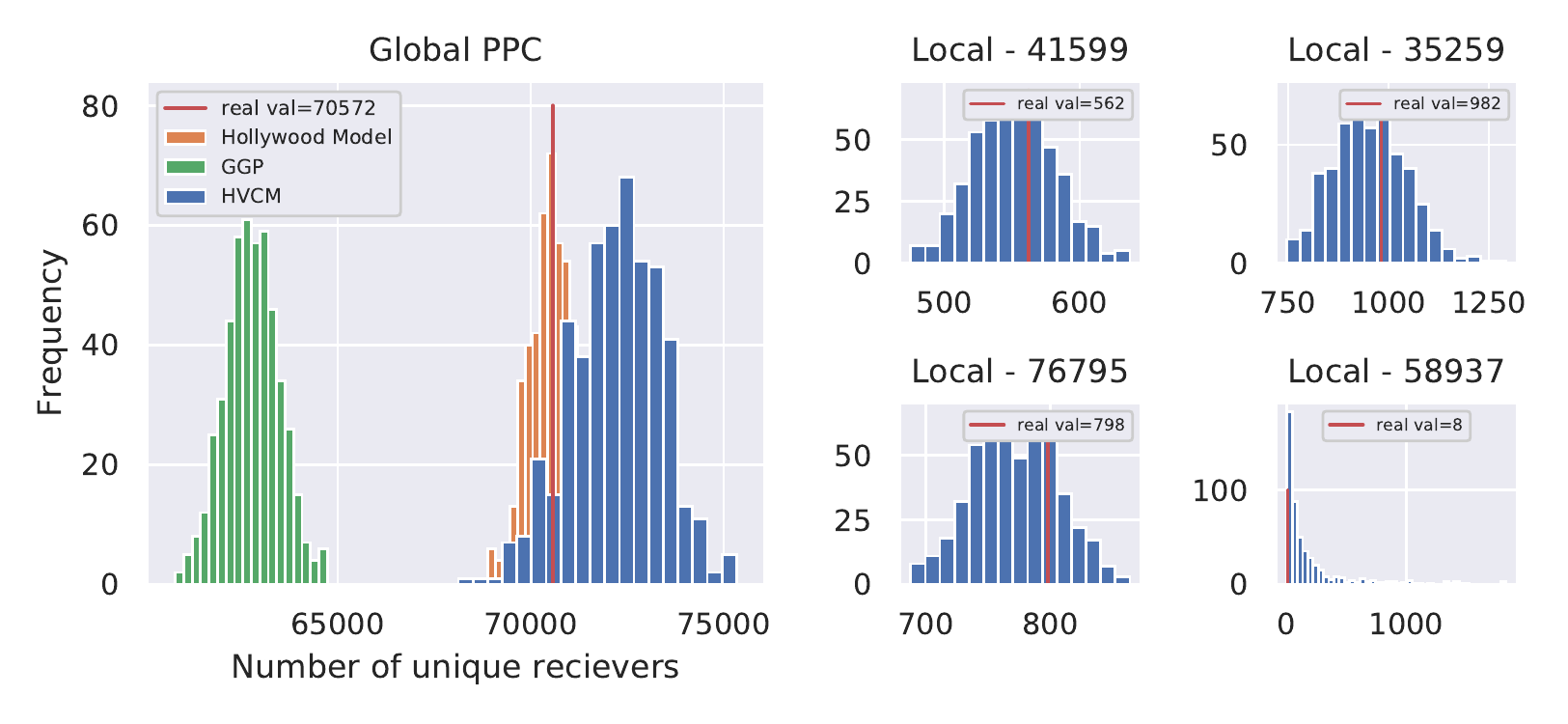}
  \caption{PPC Statistics for number of unique receivers, global (left) and
    examples of local (right).}
  \label{fig:num_verts_enron}
\end{figure}

The first statistic we consider is the number of unique receivers, both in the
global dataset as well as each local sender datasets. The number of unique
receivers can be thought of as a surrogate for sparsity, and thus an important
statistic for a candidate model to replicate. Figure~\ref{fig:num_verts_enron}
displays the results.

On the left plot, the PPC statistics are shown for the number of unique receivers
in the global dataset. Both the Hollywood model and the proposed HVCM perform
well on the global statistic. On the left plots
are four examples of the  PPC statistics for the number of unique receivers on
the local sender datasets  with the most emails. Only the results from the
proposed HVCM is shown, because neither the GGP model nor the Hollywood
model is able to take into account variation among the local distributions; if
the sender labels are attached post-hoc to the synthetic data generated from
the GGP or Hollywood model, they are completely unable to replicate any local
behavior statistics. The  HVCM clearly accounts for the varying local
behavior, even when that local behavior is unusual (in the case of
sender 58937). The superiority of the model compared to the Hollywood
model is clearly shown in Table~\ref{table:ppc_local_enron}, as the
proposed model's local posterior predictive intervals in the local
distributions covers the real values 99\% of the time, as opposed to
the Hollywood model's coverage rate of 39\%. %In order
%to create these synthetic datasets, each posterior sample of the parameters was
%used, along with the same sender sequence and size of email, to generate a
%dataset based on the Enron model. The purpose of these datasets is to compare
%them with the real data, and determine if we are retaining the important
%information of the original data.

\subsubsection{Degree distribution}
\label{subsubsection:csn_power}
\begin{figure}[ht]
\centering
\includegraphics[width=5in]{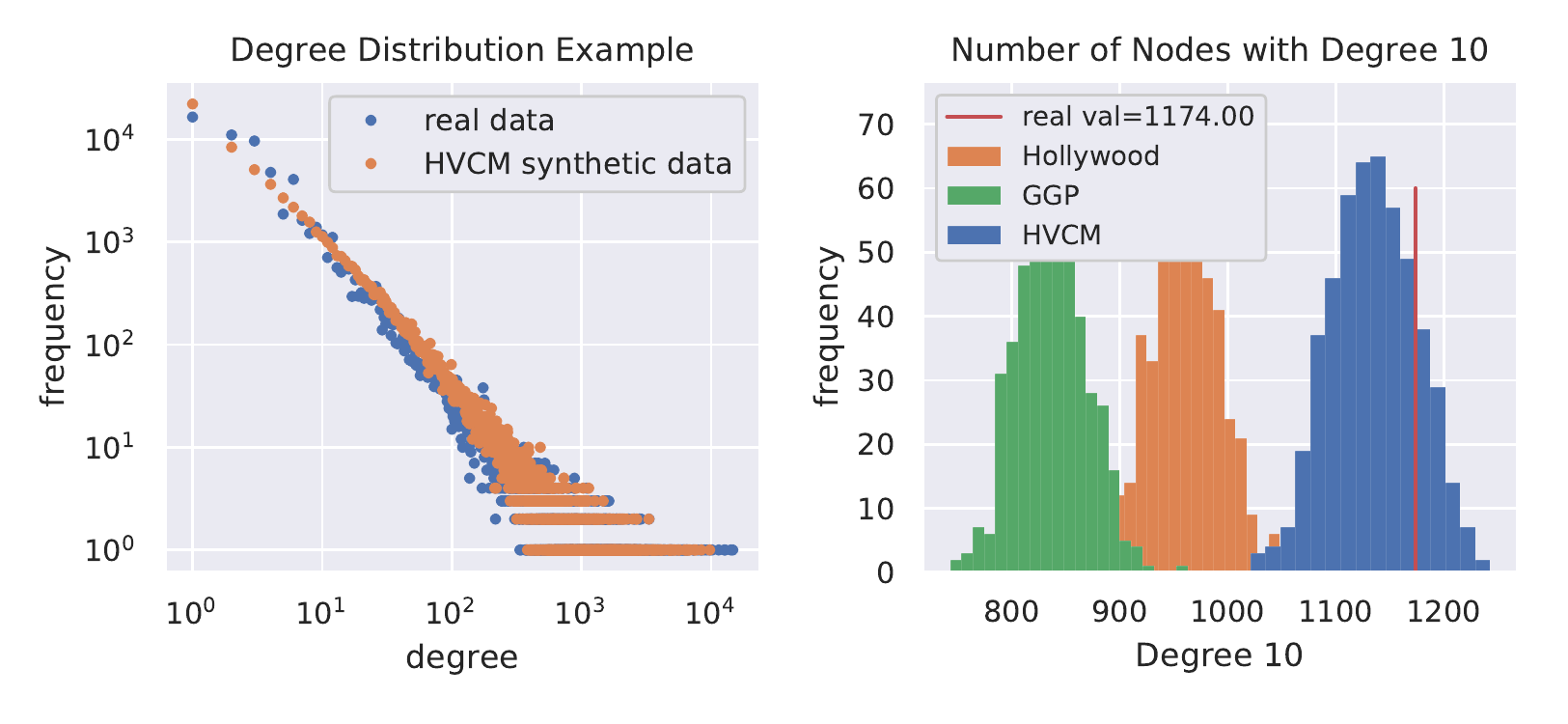}
\caption{Comparison of degree distribution between a posterior predictive sample
  from the proposed model and the real data (left) and PPC of the number of receivers with degree 10.}
\label{fig:power_ppc_enron}
\end{figure}

\begin{figure}[ht]
\centering
\includegraphics[width=5in]{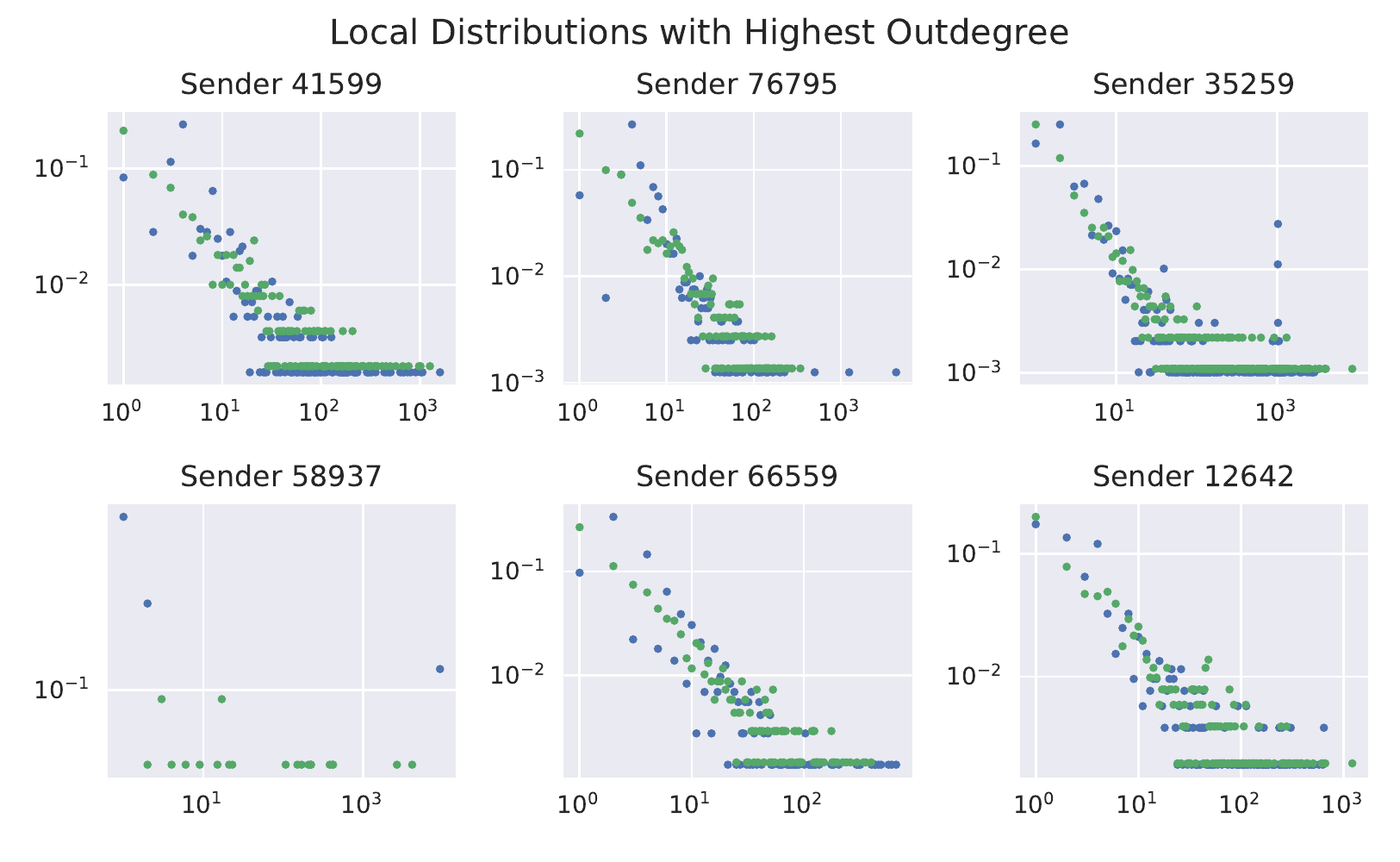}
\caption{Comparison of HVCM and Hollywood model for local distributions.}
\label{fig:enron_compare_local}
\end{figure}

An important global behavior to capture is the global degree
distribution. In order to evaluate this fit, posterior predictive
intervals of the number of nodes with degree 1, 10, and 100 are shown
in Table~\ref{table:ppc_global_enron}. Note that the HVCM performs the
best, where the real number of receivers with degree
10 are within the PP interval.~Figure~\ref{fig:power_ppc_enron} shows
this result. When comparing the degree distributions, it is also clear
that the Enron data does not perfectly align with the posterior
predictive example, as the synthetic data overestimates the number of
receivers with degree 1 and underestimates the number of receivers with
degree 100. However, it is also clear that this model fit is still
superior to both comparators, via Table~\ref{table:ppc_global_enron}. Further, 
Table~\ref{table:ppc_local_enron} demonstrates that the coverage for the posterior predictive intervals is much more robust in the proposed model for each of the degree statistics. Figure~\ref{fig:enron_compare_local} also compares local
degree distributions between the HVCM and the real data. In the both
the global and local case, the HVCM is able to better replicate the
degree distribution.

 %, and %in particular the power-law
%coefficient. In order to estimate the power law, the %Clauset-Shalizi-
%Newman (CSN) model is used, along with the appropriate %estimation
%technique~\cite{ClausetShaliziNewman2009}.  On the left is the global
%degree distribution of one posterior predictive sample laid over the
%%real degree distribution of the Enron dataset. On the right is the
%e%stimated CSN power law for the posterior predictive distribution for
%t%he Hollywood, GGP and proposed hierarchical edge exchangeable model.
%W%e note that the power-law exponent is quite sensitive, and the visual
%c%omparisons look much closer than the posterior predictive check would
%indicate. The proposed model improves over both the Hollywood and GGP
%models. Figure~\ref{fig:enron_compare_local} also compares local
%degree distributions between the hierarchical edge exchangeable model
%and the real data. In the both the global and local case, the
%hierarchical exchangeable model is able to better replicate the degree
%distribution.

%predictive sample 
%First, Figure~\ref{fig:enron_compare} compares the global degree
%distributions of a synthetic dataset instance with the real dataset,
%both for the Enron and Hollywood models. We see that the Enron model
%does a better job of modeling the beginning of the global degree
%distribution than the Hollywood model does, particularly when the
%degree in the smaller regime. We posit that this is due to the Enron
%model allowing for heterogeneity in the local sender
%distributions. This is more explicitly demonstrated when comparing the
%synthetic local degree distributions with the real data in
%Figure~\ref{fig:enron_compare_local}. 

\begin{table*}\centering

  \begin{tabular}{@{}rrrrr@{}}\toprule
    & \multirow{2}{*}{Unique Receivers} & \multirow{2}{*}{\parbox{2.5cm}{Receivers with degree 1}} & \multirow{2}{*}{\parbox{2.5cm}{Receivers with degree 10}} & \multirow{2}{*}{\parbox{2.5cm}{Receivers with degree 100}} \\ \\
\midrule HVCM & \textbf{(69881, 74299)} & (21504, 23022) &
\textbf{(1057, 1207)} &  (30, 56)  \\ Hollywood Model & \textbf{(69382, 71671)}
& (23031, 23997) & (893, 1022) &(31, 59) \\ GGP Model & (61309, 64175 )& (20653,
22006) & (778, 898) & (26, 51) \\ Actual Value & 70572& 16495 & 1174 & 15 \\
\bottomrule
\end{tabular}
\caption{Posterior predictive confidence intervals (95\%) for global statistics}
\label{table:ppc_global_enron}
\end{table*}

\begin{table*}\centering
\begin{tabular}{@{}rrrrr@{}}\toprule
 & \multirow{2}{*}{Unique Receivers} & \multirow{2}{*}{\parbox{2.5cm}{Receivers with degree 1}} & \multirow{2}{*}{\parbox{2.5cm}{Receivers with degree 10}} & \multirow{2}{*}{\parbox{2.5cm}{Receivers with degree 100}} \\ \\ \midrule
HVCM & 19725 / 19752 & 18233 / 19752 & 808 / 960 &  14 / 22  \\
Hollywood Model & 7652 / 19752 & 7652 / 19752 & 48 / 960 & 1 / 22\\
\bottomrule
\end{tabular}
\caption{Posterior predictive coverage rates of the local distributions when using the 95\% posterior predictive interval.}
\label{table:ppc_local_enron}
\end{table*}

\subsubsection{Node sharing across local distributions}

\begin{figure}[ht]
\centering
\includegraphics[width=3in]{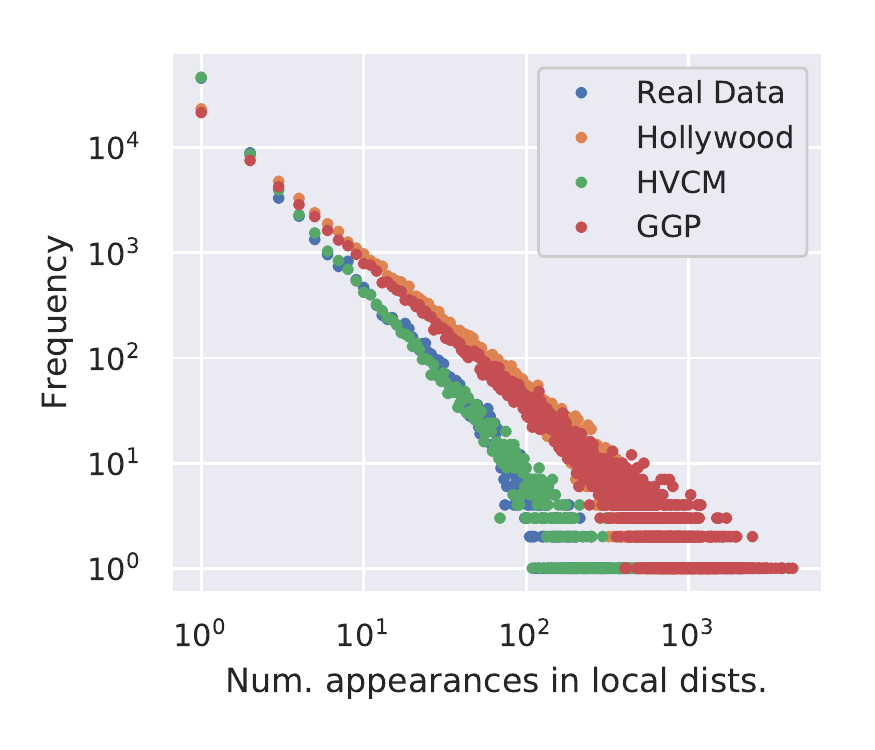}
\caption{Distribution of nodes that have been in $x$ number of local sender distributions.}
\label{fig:enron_sharing}
\end{figure}

In order to visualize how effectively the proposed HVCM is capturing the varying dependencies between the
local and global distributions, we count the number of receivers that
are seen in a particular number of local sender distributions. This
allows for direct comparison of the effectiveness of the models to
capture the interdependency and interaction among the local
datasets.~Figure~\ref{fig:enron_sharing} shows the results.

It is quite clear that the proposed HVCM replicates the observed
behavior in the real data, while both the GGP and Hollywood models
fail to capture the degree of pooling across the local
datasets. Specifically, the other models seem to overestimate the rate
at which receivers are shared across the local distributions.

\subsubsection{L1 distance from degree distribution}
\label{subsubsection:locals}

\begin{figure}[ht]
\centering
\includegraphics[width=2.5in]{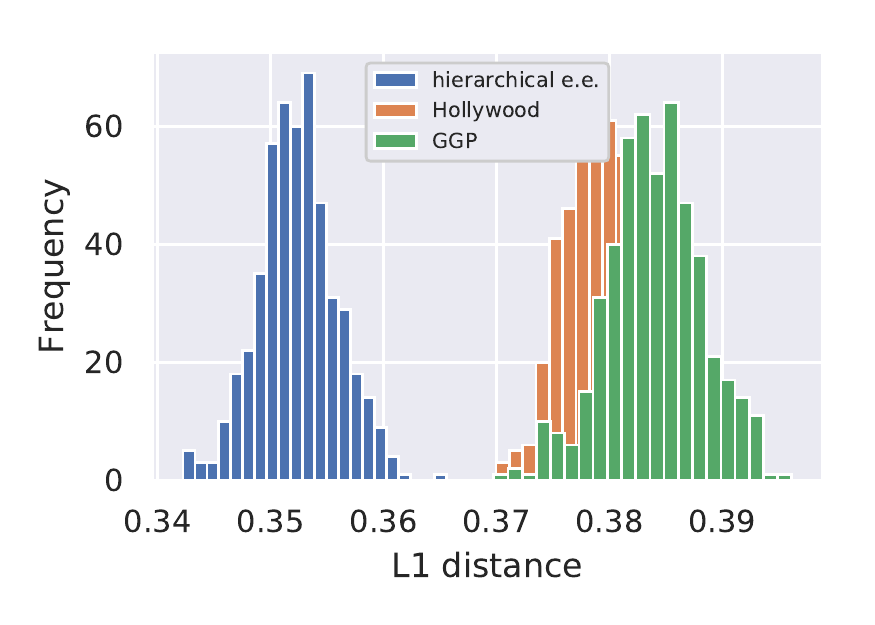}
\caption{Histograms of L1 distance between degree distributions of synthetic PPC
  datasets and Enron global degree distribution. The proposed model better
  captures the distribution than the Hollywood model and GGP model.}
\label{fig:enron_tv}
\end{figure}

With our posterior predictive samples, we can also directly examine the
difference in distribution between synthetic data and the real data.
Figure~\ref{fig:enron_tv} shows histograms of the TV distance between the global
degree distributions of the synthetic data generated from the posterior
predictive distribution and the real dataset. Again, the proposed HVCM leads to an
improvement over the Hollywood model and GGP model.
%\subsection{Prediction of growth dynamics}\label{section:finp}

% \begin{rmk}[Probability of a receiver given potential sender
%   information]
% Suppose you knew the global degree distribution.  Then this is easy
% using the setup from E2.  We then simply need to integrate
% over the unknown global degree distribution (i.e., use Gibbs samples).
% \end{rmk}

%\section{Posterior predictive checks for interaction data}

%\begin{defn}{Invariant statistic}
%Let~$y \in \Y$ then for any permutation~$\sigma :[n] \to [n]$
%\[
%y^{\sigma} = (y_{\sigma (1)}, \ldots, y_{\sigma (n)}) \in \Y.
%\]
%That is,~$y^\sigma$ is the re-ordered sequence of interactions.
%We say that~$T(y)$ is an \emph{invariance statistic} if
%$T(y) = T(y^{\sigma})$ for all
%\end{defn}

%This invariance is guaranteed for independent data, but it is 
%critical to address this issue for complex structural data such
%as interaction networks. A simple invariant statistic is 

%\section{Temporal interaction data} \label{section:temporaldata}

%\subsection{Temporal Enron process} \label{subsection:tempEnron}

%\subsection{Bayesian posterior inference} \label{subsection:inf_tempEnron}

%\subsection{Case study: Enron dataset} \label{subsection:casestudy_tempEnron}

%{\color{red} Can we find a different temporal interaction dataset?}

\section{ArXiv dataset}
\label{section:arxiv}
In this section, a larger and more complex dataset is used to demonstrate
the flexibility of the proposed HVCM. The hierarchical exchangeable model is applied to the
arXiv dataset \url{https://archive.org/details/arxiv-bulk-metadata}, which
contains nearly all arXiv articles from 1986 to 2017. Like the Enron dataset,
the arXiv data has a hierarchical structure --- each article is required to have
at least one associated subject. However, unlike the email dataset, which had
only one sender per email, each article may have more than one subject. Our
proposed model is well suited to this 
case of multiple entities 
and the data can still be appropriately represented by Equation~\ref{eq:paintbox}.
Further, our model allows for the direct study of interdisciplinarity among authors
and overlap among the subject classes on arXiv.

The arXiv subjects have been divided into 11 main classes;
the full list can be found on \url{https://arxiv.org/help/prep}.  In order to
reduce the effect of author name ambiguity, we restricted ourselves to articles
which have at least one subject from the \verb+math, cs, stat,+ and
\verb+physics+ subject classes. A full description of the subjects of interest is found in
section~\ref{appendix:arxiv} of the supplementary materials. Figure~\ref{fig:subject_hist} shows a degree
distribution for the subjects, along with a histogram of the number of subjects
per article. In total, there are 510812 scientific articles with 413029 unique authors and 130 unique subjects, 
There is also a broad range of subject frequencies, with the
most popular subject being \verb+math-ph+ (mathematical physics) with 47942 articles, and the least
popular subject \verb+cs.GL+ (general literature) with 130 articles.

\begin{figure}[ht]
  \centering
  \includegraphics[width=6.5in]{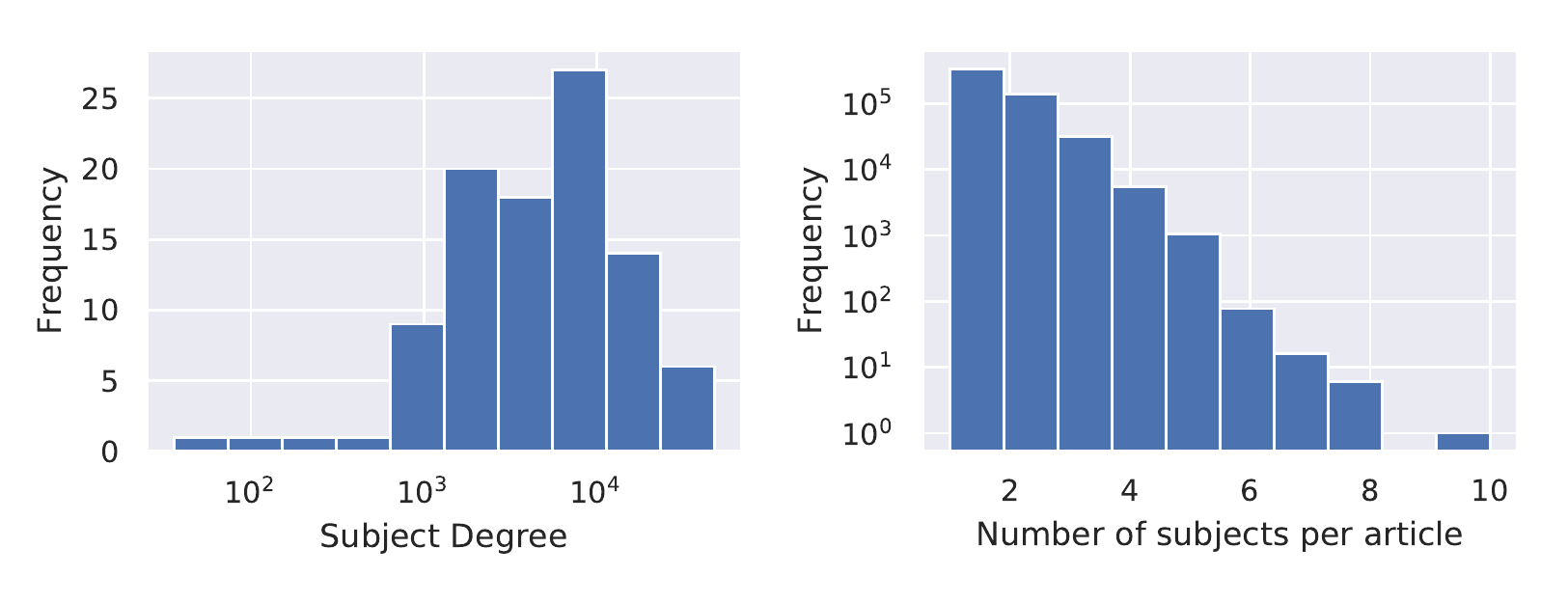} 
  \caption{Degree distribution of subjects, log scale. This degree distribution
    does not exhibit a power-law. The proposed HVCM accounts for this extra
    complexity. ($\alpha_s$ are not constrained to be equal to 1.)}
  \label{fig:subject_hist}
\end{figure}

We apply our posterior sampling methods found in
Section~\ref{section:estimation}, and in particular use the approximate method of
calculating the posteriors of the indicator variables $Z_i$ using the methods
described in~Section~\ref{subsection:approx_collapse}. Trace plots of posterior
estimates of certain parameters, posterior predictive checks for the
data and other details of the inference can be found in
section~\ref{appendix:arxiv} of the supplementary materials.

%Figure~\ref{fig:ppc_arxiv} shows PPC statistics for two
%global properties: number of total unique authors and estimated CSN power
%exponent. We see that the estimated power law exponent for both sets of
%synthetic data overlaps the estimated power law exponent for the arXiv %dataset. Moreover, the model performs
%well in the number of vertices.

%\begin{figure}[ht]
%  \centering
% \includegraphics[width=\textwidth]{./figs/ppc_arxiv.eps} 
%  \caption{Posterior predictive check for number of unique authors and power law
%    fit for arXiv data.}
%  \label{fig:ppc_arxiv}
%\end{figure}

\subsection{Subject Overlap}
\label{ssec:subject_overlap}
The fitted model allows us to explore the amount of overlap between arXiv
subjects. Two subjects are considered overlapping if the model has difficulty
distinguishing between them when they are used as labels for the same article.
This difficulty can be measured using the Shannon entropy, which is defined over
discrete probability distributions $p = [p_1, p_2, \ldots, p_k]$ as:

$$
H(p_1, p_2, \ldots, p_k) = -\sum_{k} p_k\log_2 p_k.
$$

Entropy is at its maximum when the distribution $p$ is the uniform distribution,
i.e., when all outcomes are equally likely. In order to estimate subject
overlap for subjects $s_1$ and $s_2$, every article which lists $s_1$ and $s_2$
among its subjects is found, and the entropy of the posterior mean of the $Z_i$
distribution given that the subject is either $s_1$ or $s_2$ is calculated, and
the entropy is averaged over the articles. This score, $\text{SO}(s_1, s_2)$ is
computed as:

\begin{equation}
  \label{eq:subject_overlap}
\text{SO}(s_1, s_2) = \frac{1}{|\{\bar S_i : s_1, s_2 \in \bar S_i\}|} \sum_{i : s_1,
  s_2 \in \bar S_i} H(\pr(Z_i = s | \{\bar S_i, \bar R_i \}, Z_i \in \{s_1, s_2\}))
\end{equation}

\begin{figure}[ht]
  \centering
 \includegraphics[width=\textwidth]{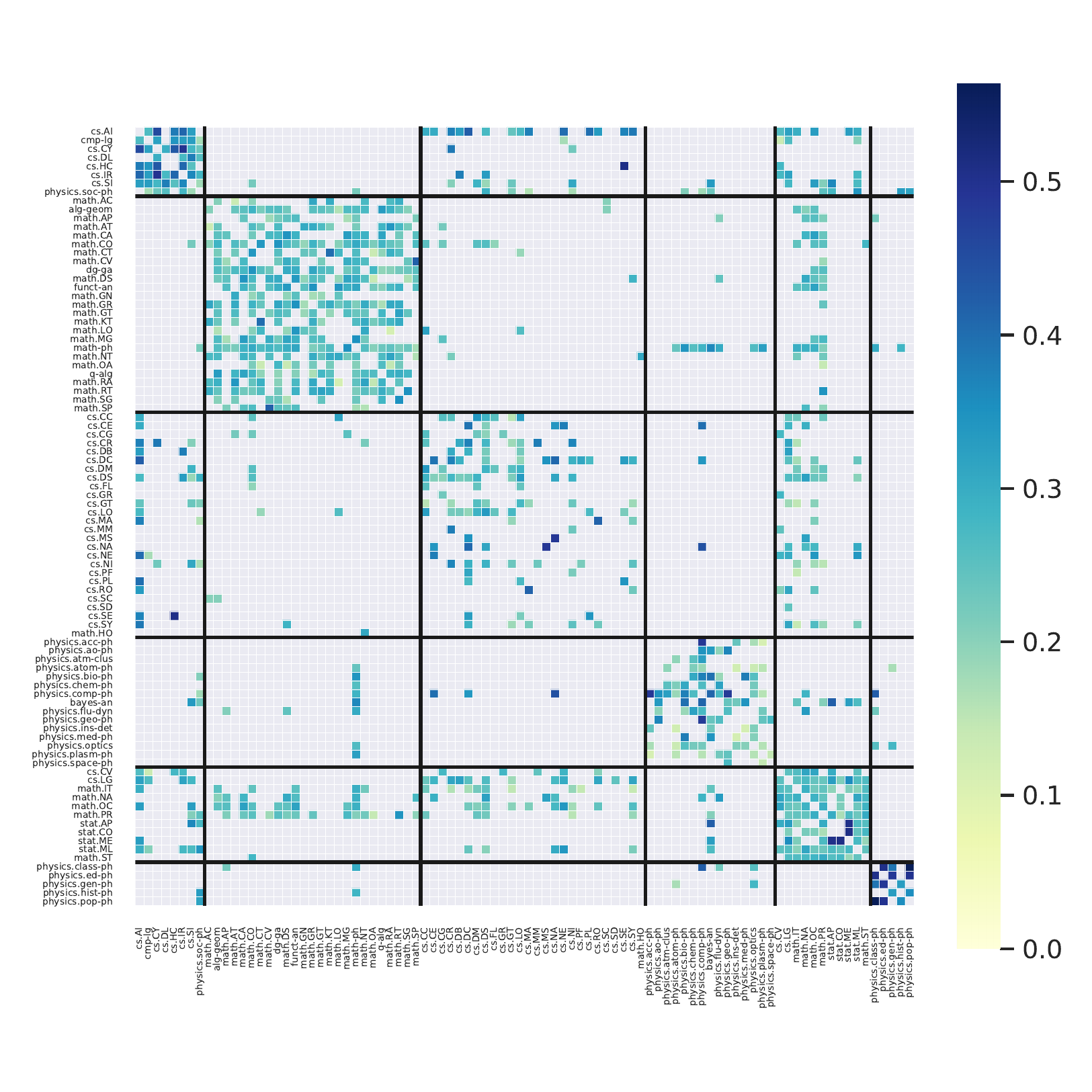} 
  \caption{Heatmap of two-way entropy per article. For each pair of subjects
    $s_1, s_2$, and every article that contains both $s_1$ and $s_2$, the
    entropy of $\pr(Z_i = z| Z_i \in \{s_1, s_2\})$ is calculated and summed.
    Finally, each entry is normalized by the total number of occurences of $s_1$
    and $s_2$ appearing together in the same article. }
  \label{fig:sub_overlap}
\end{figure}

Figure~\ref{fig:sub_overlap} shows a heatmap of the subject overlap scores for
subjects that are seen in the same article at least 100 times. The subjects are
ordered according to a normalized spectral clustering~\cite{ng2002spectral},
using the subject overlap matrix $\text{SO}$ as
the affinity matrix, and setting the number of clusters to 6.

From this analysis, we conclude the following.
Cluster 1, which includes cs.AI (Artificial
Intelligence) and cs.IR (Information Retrieval), is a group of subjects that
pertain to algorithmic approaches to artificial
intelligence. Note that this cluster is differentiated from cluster 5, which 
tends to represent more theoretical papers that use statistics;
this cluster includes math.ST
(Statistical Theory), stat.ML (Machine Learning), and stat.ME (Methods). Cluster
2 can be considered the core math cluster, which encapsulates many pure and applied
math subjects. Similarly, cluster 3 is the core computer science cluster, which
are the computer science subjects that generally don't use statistics
such as cs.SE (Software Engineering) and cs.CE (Computer
Engineering). Cluster 4 is the core physics cluster, with the subjects of
physics that tend not to be interdisciplinary outside of physics as other
physics subjects. Finally,
cluster 6 consists of subjects that involve the philosophy, teaching or history
of physics. Perhaps unsurprisingly, as AI is a fast moving field, two clusters are found (clusters 1 and
4) within AI that do not have a designated arXiv category.
Table~\ref{table:top3}
lists the pairs of subjects with the most overlap according to the entropy score~\ref{eq:subject_overlap}. Note that these pairs
correspond with the general intuition of subjects that would have a large degree
of interdisciplinarity.
%Table~\ref{table:top3} shows the top 3 subject overlap scores with the
%corresponding subject pairs.

\begin{table}[ht]
  \caption{Pairs of subjects with highest subject overlap score.}
  \centering
\begin{tabular}{ |c|c|c| }
 \hline
$s_1$ & $s_2$ & $\text{SO}(s_1, s_2)$ \\ \hline
  stat.ME (Methods)      & stat.CO (Computation) & 0.509 \\
  cs.SE (Software Engineering)      & cs.HC (High Perf. Comp.) & 0.507 \\
  physics.class-ph (Classical Physics)      & physics.ed-ph (Education)  & 0.504 \\
  \hline
\end{tabular}
%\caption{Pairs of subjects with highest overlap score.}
\label{table:top3}
\end{table}

We compare these results with results of a direct application of a spectral clustering
algorithm to the co-authorship
network in section~\ref{appendix:arxiv} of the supplementary
materials. This direct application of spectral clustering to the data
is unable to recover the meaningful groupings that the proposed HVCM produces. 
%\textcolor{red}{I'd add a few sentences on the inability of direct
%  application of spectral clustering to pick up these clusters.
%  I'd also say something about comparison to hollywood and GGP not possible due
%  to the complex structure.}
\section{Concluding remarks}
\label{section:discuss}
This paper has presented the class of exchangeable structured interaction models.  By
exploiting the common hierarchical nature of structured network data,
complex models with both appropriate invariance and empirical
properties are introduced.
The canonical HVCM captures partial pooling of information, and can
model complex local-behavior with global power-law degree behavior. 
A Gibbs sampling algorithm is proposed and applied to the Enron e-mail
and arXiv datasets. 
The focus of this paper has been on e-mail and similarly structured
interaction datasets. Extensions to more complex examples will be in 
considered future work. This paper lays the foundation for
how the interaction exchangeability framework can account for complex
behavior.
Of course, many interaction networks occur with time-stamps;
therefore, extensions to account for temporal dependence is required
and will be an important next step .

\clearpage
\bibliographystyle{unsrt}
\bibliography{network-refs}

\end{document}